\documentclass{article}
\usepackage[a4paper,top=3cm,bottom=3cm,left=3cm,right=3cm]{geometry}
\usepackage{graphicx} 
\usepackage{amsmath}
\usepackage{amssymb}
\usepackage{amsthm}
\usepackage{booktabs}
\usepackage{xcolor}
\usepackage{authblk}
\usepackage{comment}
\title{From Brain Scans to Therapy Response: PDE Modelling of Immunotherapy for Glioblastoma}

\author[1]{Francesca Ballatore\thanks{Corresponding author. Email: francesca.ballatore@univ-cotedazur.fr}}
\author[2]{Lorenzo Scolaris\thanks{Email: scolaris.lorenzo@gmail.com}}
\author[2]{Chiara Giverso\thanks{Email: chiara.giverso@polito.it}}

\affil[1]{\normalsize Laboratoire J. A. Dieudonné, Université Côte d’Azur, Nice, France}
\affil[2]{\normalsize Dipartimento di Scienze Matematiche ``G.L. Lagrange'', Politecnico di Torino, Italy}

\date{ }
\usepackage[backend=biber,style=ext-numeric-comp,autocite=inline, sorting=none, giveninits=true, articlein=false, date=year, maxbibnames=500]{biblatex}
\begin{document}

\maketitle

\section*{Abstract}
Glioblastoma Multiforme (GBM) is a highly aggressive brain tumour with limited therapeutic options and poor prognosis. This study presents a mathematical framework to investigate the efficacy of immunotherapy strategies based on cytotoxic T-lymphocyte (CTL) infusion. The model couples tumour and immune dynamics through a system of partial differential equations (PDEs), incorporating cell proliferation, diffusion, and chemotactic migration in response to TGF-$\beta$, a tumour-secreted signalling molecule. A reduced ordinary differential equation (ODE) model is first analysed to derive threshold conditions for tumour eradication, identifying critical infusion levels consistent with clinical data. Numerical bifurcation analysis explores the impact of parameter variations.
The full PDE model is solved using the finite element method on simplified 2D domains, followed by sensitivity analyses to quantify parameter influence on tumour mass and volume. The model is then applied to a realistic 3D brain geometry reconstructed from patient-specific MRI and DTI data, accounting for anatomical anisotropy and tissue heterogeneity. Therapeutic scenarios are simulated with spatially localised lymphocyte infusion. Results highlight spatial variations in tumour growth and treatment response, with infusion intensity and tumour location critically influencing therapeutic outcomes. These findings emphasise the importance of personalised, spatially informed modelling in optimising immunotherapy protocols for GBM. 

\vspace{3mm}
\noindent\textbf{Keywords:}
Tumour–immune interactions; immunotherapy modelling; partial differential equations; brain tumors modelling

\section{Introduction}
Immunotherapy aims to generate a specific immune response that targets cancer cells, countering tumour growth while sparing the surrounding healthy tissue \cite{Pinheiro:2023}. Despite its limited use compared to conventional treatments, immunotherapy has been approved for various types of cancer, including melanoma and prostate cancer, and has sparked interest as a potential approach for Glioblastoma Multiforme (GBM). However, its application in GBM faces numerous challenges, such as the immunosuppressive tumour microenvironment, the presence of the blood-brain barrier (BBB) that grants the brain immune privilege, the intratumoural heterogeneity of GBM, and the existence of therapy-resistant cancer stem cells \cite{Agosti:2023}. These obstacles have led to the development and testing of various immunotherapy strategies, primarily focusing on vaccines, immune checkpoint inhibitors, chimeric antigen receptor T cells (CAR-T), and oncolytic viruses \cite{Maggs:2021}. 

Cancer vaccines aim to enhance immune surveillance by leveraging tumour-derived antigens \cite{Rong:2022}, but their effectiveness for GBM is constrained by the rarity and variability of specific antigens. For instance, targeting the EGFRvIII antigen, present in approximately 30\% of GBM cases, has shown promise \cite{Weller:2017, Lim:2018}. Recent advances in sequencing technologies have further identified neoantigens arising from somatic mutations, providing a highly specific basis for therapy. Similarly, immune checkpoint inhibitors have demonstrated the ability to counteract the suppressive tumour microenvironment by blocking receptors, such as PD-1 and CTLA-4, thereby restoring immune function \cite{Wainwright:2014, Zeng:2013}. Meanwhile, oncolytic viruses have emerged as a novel approach, selectively targeting cancer cells while activating the immune response through molecular patterns released during infection \cite{Iorgulescu:2018, Martikainen:2019}. CAR-T cell therapy, which involves engineering T cells to recognise tumour-specific antigens, also holds significant potential, though its efficacy against GBM is hindered by the BBB and the tumour's immunosuppressive environment \cite{Choi:2018}. 

The emergence of immunotherapy has spurred the development of mathematical models to describe the complex interactions between tumours and the immune system. These models typically focus on tumour cells and immunocytes, such as cytotoxic T lymphocytes (CTLs) and macrophages, and often include mediators like cytokines and antigens \cite{Konstorum:2017, Tang:2023}. Early approaches employed predator-prey ODE systems, such as those by Kuznetsov \cite{Kuznetsov:1994} and Kirschner and Panetta \cite{Kirschner:1998}, to capture tumour recurrence and immune cell overgrowth. More recent models incorporate spatial effects using reaction-diffusion PDEs \cite{Lazebnik:2022} or hybrid frameworks, integrating cellular dynamics with chemical diffusion \cite{Wells:2015}. Brain tumour-specific models include ODE frameworks that simulate glioblastoma interactions with CTLs \cite{Kronik:2008}, and extensions with delays or spatio-temporal dynamics for capturing therapeutic responses \cite{Khajanchi:2018, Khajanchi:2021}.

In this work, an advection-diffusion-reaction model is proposed to characterise the immunotherapy response in tumours growing within highly anisotropic environments, such as glioblastoma. The model is designed to capture tumour cell proliferation throughout disease progression and its interactions with the immune system, including the potential effects of immunotherapy through cytotoxic T lymphocyte infusions. 
Section \ref{sec:mathematical_model} outlines the mathematical formulation adopted in this study. Section \ref{sec:qualitative_analysis} introduces a reduced ODE-based version of the model, which describes tumour proliferation and the impact of CTL infusion on the malignant cell population. The equilibria of the system and the influence of therapy on the trajectories are investigated both analytically and numerically. Conditions for therapy effectiveness are derived, and bifurcation analyses are performed on key model parameters. 
Finally, Section \ref{sec:numerical_simulations} presents the results of numerical simulations. The model is first solved in a simplified 2D geometry to verify code stability and assess system behaviour, followed by a sensitivity analysis to quantify the relative influence of each parameter on tumour treatment progression. The model is then applied to a realistic 3D brain geometry obtained from magnetic resonance imaging (MRI) scans, incorporating diffusion tensors derived from diffusion tensor imaging (DTI) to account for the anisotropic migration of cells along brain tissue fibres.

\section{Mathematical model}
\label{sec:mathematical_model}
To qualitatively capture the growth dynamics of malignant gliomas and the impact of immunotherapeutic treatment, specifically T-cell infusions, we develop a mathematical model grounded in biological and clinical observations.
An earlier mathematical model by Banerjee et al. \cite{Banerjee:2015} described glioma cell proliferation and their interactions with the immune system through a system of coupled nonlinear ordinary differential equations, incorporating the effects of T11 Target Structure, an immunotherapeutic agent. While the model offered valuable insights, its scope was limited, as it did not consider malignant glioma cell migration or the dynamic nature of immune responses.
Indeed, the migration of glioma cells within the brain is influenced by multiple factors. In clinical settings, glioma cells are frequently observed to migrate along preferential anatomical pathways, such as white matter tracts and the basal lamina of cerebral blood vessels. Their motility is further enhanced by the secretion of signalling molecules that induce chemotactic responses. These behaviours highlight the importance of specific substrates and brain microstructures in guiding glioma dispersion. Ultimately, the extent and pattern of tumour invasion result from a complex interplay between the intrinsic properties of glioma cells and the features of their surrounding microenvironment.

We model the pointwise concentrations of tumour cells and lymphocytes, denoted by $G(\boldsymbol{x},t)$ and $C(\boldsymbol{x},t)$, respectively, as functions of space and time. For both cell populations, the model incorporates diffusion, which is generally anisotropic due to the underlying tissue structure.
Additionally, lymphocyte dynamics include chemotactic movement directed toward regions with higher concentrations of transforming growth factor beta (TGF-$\beta$) which is secreted by tumour cells.
The evolution of the diffusing chemoattractant TGF-$\beta$, denoted in the model by $T(\boldsymbol{x},t)$, is described by an additional equation. The complete model is, thus, expressed as follows
\begin{equation}
\begin{cases}
\label{dimentional_PDE_model}
& \dfrac{\partial G}{\partial t}= \nabla \cdot \left( \mathbb{D}_G \nabla G \right)+r G \left(1-\dfrac{G}{G_m} \right)-\alpha_G C \dfrac{G}{k_G+G} \, , \vspace{3mm}\\ 
& \dfrac{\partial C}{\partial t}=\nabla \cdot \left( \mathbb{D}_C \nabla C \right)-\nabla \cdot[\chi(G,C,T) C \nabla T]+b_C-\mu_C C-\alpha_C C \dfrac{G}{k_C+G}+S_T(\boldsymbol{x}, t) \, ,\vspace{3mm}\\
& \dfrac{\partial T}{\partial t}=\nabla \cdot \left( \mathbb{D}_T \nabla T \right)+p_T G-\mu_T T \, .
\end{cases}
\end{equation}
The first term on the right-hand side of the first equation of system \eqref{dimentional_PDE_model} represents the random motility of malignant glioma cells, where $\mathbb{D}_G$ is the diffusion tensor. The second term models the growth of malignant gliomas using a logistic growth function, with $r$ denoting the proliferation rate and $G_{m}$ the tumour carrying capacity. The final term accounts for the eradication of glioma cells by cytotoxic T-lymphocytes, occurring at a rate $\alpha_G$, following Michaelis-Menten kinetics, where $k_G$ is the half-saturation constant associated with tumour cell death induced by the immune response.

The second equation of system~\eqref{dimentional_PDE_model} characterises the dynamics of activated cytotoxic T-lymphocytes, incorporating both their migration and diffusion mechanisms.
The first term on the right-hand side models diffusion, where $\mathbb{D}_C$ is the diffusion tensor. The second term represents chemotaxis, capturing the directed movement of cytotoxic T-lymphocytes toward regions with higher concentrations of TGF-$\beta$. The chemotactic sensitivity coefficient, denoted by $\chi$, is assumed, in general, to be a function of the tumour cell concentration, the T-cell concentration, and the local TGF-$\beta$ concentration.
Cytotoxic T lymphocytes are recruited and activated by the immune system at a constant rate $b_C$, while their natural decay occurs at a rate $\mu_C$.
The clearance of cytotoxic T-lymphocytes by malignant glioma cells follows Michaelis-Menten saturation kinetics, with rate $\alpha_C$ and half-saturation constant $k_C$. This formulation captures the saturation effects arising in the immune response due to the presence of tumour cells.
Finally, we account for the infusion of T-cells as part of the immunotherapeutic treatment, represented by the term $S_T(\boldsymbol{x},t)$.

The last equation of system \eqref{dimentional_PDE_model} describes the dynamics of the chemoattractant TGF-$\beta$. Glioma cells release TGF-$\beta$ as part of their interaction with the tumour microenvironment, with the amount secreted increasing proportionally to the tumour cell population. For simplicity, a linear relationship, with constant production rate $p_T$, between TGF-$\beta$ secretion and tumour cell density is assumed in this model, acknowledging that the underlying biological mechanisms are far more complex and may involve non-linear interactions and context-dependent regulatory effects. We assume that TGF-$\beta$ diffuses throughout the tissue with rate dictated by the diffusion tensor $\mathbb{D}_T$. TGF-$\beta$ is also supposed to decay linearly at the rate $\mu_T$. 

Concerning the boundary conditions, homogeneous Neumann conditions are imposed on all variables, implying zero flux across the boundary of the domain for each quantity:
\begin{equation}
\nabla G \cdot \boldsymbol{n}=0, \quad \nabla C \cdot \boldsymbol{n}=0, \quad \nabla T \cdot \boldsymbol{n}=0 \, , \quad \text{on} \ \partial \Omega \, , \ t > 0 \, ,
\end{equation}
where $\partial \Omega$ represents the boundary of a simply connected domain, $\Omega$.

For the initial conditions, we assume that the initial concentration of each quantity is known and, in general, may depend on space:
\begin{equation}
G(\boldsymbol{x},0)=G_0(\boldsymbol{x})>0 \, , \ C(\boldsymbol{x}, 0)=C_0(\boldsymbol{x})>0\, , \ T(\boldsymbol{x}, 0)=T_0(\boldsymbol{x})>0 \, , \ \boldsymbol{x} \in \Omega \, .
\end{equation}

\subsection{Parameter selection}
\label{Parameter_List}
Identifying and accurately estimating the numerical parameters is essential for developing robust and reliable predictive models. However, for most of the quantities involved, such parameters are challenging to obtain through direct measurements in vivo and must be inferred from in vitro experiments or estimations. In this work, all parameters have been selected based on previously published experimental and modelling studies, as discussed in this section. 
A summary of the parameter values used in system~\eqref{dimentional_PDE_model} is presented in Table~\ref{tab:parameters}, with a distinction between the values adopted for both ODE and PDE systems.

\begin{table}[ht]
\centering
\begin{tabular}{cccc}
\toprule
\textbf{Parameter} & \textbf{Value (ODE setting)} & \textbf{Value (2D-3D settings)} &  \textbf{Refs.} \\ \midrule
$G_{m}$ & 882650 cells & $2.39 \cdot 10^8$ cells/mL & \cite{Banerjee:2015}\\
$r$ & $0.003125 \mathrm{~h}^{-1}$ & $0.003125 \mathrm{~h}^{-1}$ & \cite{Swanson:2002} \\
$\alpha_G$ & 0.12 $\mathrm{h}^{-1}$ & 0.12 $\mathrm{h}^{-1}$ & \cite{Banerjee:2015}\\
$k_G$ & 0.031$G_{m}$ & $7.409 \cdot 10^6$ cells/mL& \cite{Banerjee:2015} \\
$\alpha_C$ & 0.168 $\mathrm{h}^{-1}$ & 0.168 $\mathrm{h}^{-1}$& \cite{Banerjee:2015} \\
$k_C$ & 0.379$G_{m}$ & $9.0581 \cdot 10^7$ cells/mL& \cite{Banerjee:2015} \\
$\mu_C$ & $0.0074 \mathrm{~h}^{-1}$ & $0.0074 \mathrm{~h}^{-1}$ & \cite{Banerjee:2015} \\
$b_C$ & 74 cells $\mathrm{h}^{-1}$ & 1.85 cells/(mL $\cdot$ h) & \cite{Evans:2019}\\
$\mu_T$ & 0.1022 $\mathrm{h}^{-1}$ & 0.1022 $\mathrm{h}^{-1}$ & \cite{Banerjee:2015} \\
$p_T$ & $5.7\cdot 10^{-6}$ $\mathrm{pg}/(\mathrm{cells} \cdot \mathrm{h})$ &$5.7\cdot 10^{-6}$ $\mathrm{pg}/(\mathrm{cells} \cdot \mathrm{h})$ & \cite{Banerjee:2015} \\
$D_G$ & - & $5.417 \cdot 10^{-5} \, \mathrm{cm}^2/ \mathrm{h}$  & \cite{Swanson:2002} \\
$D_C$ & - & $4.167 \cdot 10^{-4} \, \mathrm{cm}^2/ \mathrm{h}$ & \cite{Matzavinos:2004} \\
$D_T$ & - & $6.6336 \cdot 10^{-3} \, \mathrm{cm}^2/ \mathrm{h}$ & \cite{Matzavinos:2004, Khajanchi:2021}\\
\bottomrule
\end{tabular}
\caption{List of dimensional parameters used in stability analysis.}
\label{tab:parameters}
\end{table}

Regarding the equation governing tumour cell dynamics in the brain, the carrying capacity is set at \( G_{m} = 882650 \) cells, calibrated using growth curves from \emph{in vitro} experiments \cite{Banerjee:2015}. In that study, the spatial distribution of cells is not considered; therefore, this value is appropriate for use in the ODE system, which describes tumour progression over time. For three-dimensional simulations within brain geometry, where the focus lies on local cell concentration, we adopt a carrying capacity of \( G_{m} = 2.39 \cdot 10^8 \) cells/mL, assuming an average cellular radius of \( 10 \, \mu\text{m} \) \cite{Gu:2011}. The same value is assumed for the two-dimensional setting, by hypothesising a representative layer thickness of 1 cm thereby maintaining consistency in the volumetric cell density approximation. Furthermore, for the proliferation rate, we consider \( r = 0.003125 \, \text{h}^{-1} \), as reported in \cite{Swanson:2002}.

The tumour cell elimination coefficient is taken from \cite{Banerjee:2015}, with \( \alpha_G = 0.12 \ \text{h}^{-1} \), neglecting the immunosuppressive effects of TGF-$\beta$. The corresponding Michaelis-Menten half-saturation constant is set to \( k_G = 0.031 \, G_{m} \), based on in vitro experiments reported in \cite{Banerjee:2015}.

Similarly, the lymphocyte elimination coefficient is set to \( \alpha_C = 0.168 \ \text{h}^{-1} \), while the associated Michaelis-Menten half-saturation constant is chosen as \( k_C = 0.379 \, G_{m} \), also derived from in vitro data in \cite{Banerjee:2015}. Moreover, the half-life of lymphocytes is approximately 3.9 days \cite{Banerjee:2015}, corresponding to a natural death rate of \( \mu_C = 0.0074 \, \text{h}^{-1} \). 

The central nervous system (CNS) of a healthy individual contains approximately \( 1.5 \cdot 10^5 \) lymphocytes~\cite{Evans:2019}. In the presence of inflammation, this number is hypothesised to double as a result of the innate immune response. To maintain a stable lymphocyte count in the absence of tumour-induced or therapy-induced elimination, the production rate is estimated accordingly. 
Specifically, in the one-dimensional setting, where only the portion of tissue affected by the tumour is considered, we account solely for the lymphocytes actively interacting with the tumour microenvironment. Accordingly, a reference population of \( 10^4 \) cells is assumed. Based on this value, we set the lymphocyte influx rate to \( b_C = 10^4 \cdot \mu_C = 74 \, \text{cells} \cdot \text{h}^{-1} \).
Conversely, in the 2D and 3D settings, where lymphocyte concentration is modelled, the total number of cells $3.0 \cdot 10^5$  is divided by an estimated brain volume of 1200 mL, yielding a density of approximately 250 cells/mL. Consequently, the lymphocyte source term is set to $b_C = 250 \, \text{cells}/\text{mL} \cdot \mu_C \approx 1.85 \, \text{cells}/(\text{mL} \cdot \text{h})$.

In this work, we set the natural decay rate of TGF-\(\beta\) to \(\mu_T = 0.1022\, \text{h}^{-1}\), corresponding to a half-life of approximately 6.8 hours. This value is compatible with the reported persistence of latent or matrix-bound TGF-\(\beta\) in tissue environments, which can range from few hours to over a day \cite{Wakefield:1991, Hara:2015}. Additionally, the release rate per glioma cell, \(p_T\), is also taken from \cite{Banerjee:2015}, with an assumed value of \(p_T = 5.7 \cdot 10^{-6} \, \mathrm{pg}/(\mathrm{cell} \cdot \mathrm{h})\).

The mean diffusivity of tumour cells, lymphocytes, and cytokines is defined as one-third of the trace of their respective diffusion tensors, \(\mathbb{D}_G\), \(\mathbb{D}_C\), and \(\mathbb{D}_T\). A mean diffusivity of \( D_G = 5.417 \cdot 10^{-5} \, \mathrm{cm}^2/\mathrm{h} \) is assigned to tumour cells, derived from data in \cite{Swanson:2002}. For lymphocytes, \emph{in vitro} experiments have shown a diffusivity of up to \( 10^{-2} \, \mathrm{cm}^2 /\mathrm{day} \) in the presence of a tumour \cite{Matzavinos:2004}, resulting in a selected value of \( D_C = 4.167 \cdot 10^{-4} \, \mathrm{cm}^2/\mathrm{h} \). In contrast, cytokines exhibit mean diffusivities several orders of magnitude higher than those of cells \cite{Matzavinos:2004}. Based on estimates from \cite{Khajanchi:2021}, the cytokine mean diffusivity is approximated to be 16 times greater than that of lymphocytes.

The values for the chemotactic coefficient in the simplified 2D geometry are not readily available in the literature and will therefore be estimated through numerical simulations, as detailed in Section \ref{chemio_2D}. Similarly, for the three-dimensional brain geometry, the value for the chemotactic coefficient used in simulations will be estimated in Section \ref{3D_setup}.

\section{Spatially homogeneous steady state configurations and stability}
\label{sec:qualitative_analysis}
To gain insight into the qualitative behaviour of the proposed system, we analyse in this section the spatially homogeneous steady states of the glioma–cytotoxic T lymphocyte interaction model, corresponding to the first two equations of system~\eqref{dimentional_PDE_model}, along with their stability.
To this end, we neglect both diffusion and chemotaxis, so that the equation governing the evolution of TGF-$\beta$ is not required, and the model reduces to the following system of ordinary differential equations:
\begin{equation}
\label{ODE_system}
\begin{cases}
& \dfrac{d G}{d t}= r G \left(1-\dfrac{G}{G_{m}} \right)-\alpha_G C \dfrac{G}{k_G+G} \, , \vspace{2mm}\\
& \dfrac{d C}{d t}=b_C-\mu_C C-\alpha_C C \dfrac{G}{k_C+G}+S_T(t) \, .
\end{cases}
\end{equation}
The system is first non-dimensionalised by introducing the following transformations:
\begin{equation}
G=\tilde{G}G_{m} \, , \ C=\tilde{C}C_{m} \, , \ t = \tilde{t} \tau \, ,
\end{equation}
where the tumour cell population is normalised by its carrying capacity \( G_{m} \), and the other variables are rescaled by defining
\begin{equation}
C_{m} :=\dfrac{r G_{m}}{\alpha_G} \, , \ \tau :=\frac{1}{r} \, .
\end{equation}
 We then introduce the total production rate of cytotoxic T-lymphocytes, defined as $S(t) := S_T(t) + b_C$, in order to express the system in dimensionless form:
\begin{equation}
\label{ODE_adim}
\begin{cases}
& \dfrac{d \tilde{G}}{d \tilde{t}}=\tilde{G}(1-\tilde{G})-\tilde{C} \dfrac{\tilde{G}}{\tilde{k}_G+\tilde{G}}=f(G,C) \, , \vspace{2mm}\\
& \dfrac{d \tilde{C}}{d \tilde{t}}=\tilde{S}(\tilde{t})-\tilde{\mu_C} \tilde{C}-\tilde{\alpha_C} \tilde{C} \dfrac{\tilde{G}}{\tilde{k}_C+\tilde{G}}=h(G,C) \, ,
\end{cases}
\end{equation}
where $\tilde{\alpha}_C=\alpha_C/r$, $\tilde{k}_G=k_G/G_{m}$, $\tilde{k}_C=k_C/G_{m}$, $\tilde{\mu}_C=\mu_C/r$ e $\tilde{S}(\tilde{t})=S(t)\alpha_G/(r^2G_{m})$.
For simplicity, the tildes are omitted from $G$, $C$ and $t$ in the subsequent equations.

In this analysis, we consider the case of a constant therapy rate, i.e., $\tilde{S}(\tilde{t}) \equiv \tilde{S}$, and we examine the equilibria  $(G^* \, , \, C^*)$ of the system. 
First, we observe that the value of $C^*$ can be explicitly expressed as a function of $G^*$ and $\tilde{S}$:
\begin{equation}
\label{C_star}
C^*=\dfrac{\tilde{S}}{\tilde{\mu}_C+\tilde{\alpha}_C \dfrac{G^*}{\tilde{k}_C + G^*}} \, .
\end{equation}
By substituting into the equation for \( G \), we obtain a fourth-degree polynomial in the unknown \( G^* \):
\begin{equation}
\label{poly_3}
\begin{aligned}
G^*  \Big \{  (\tilde{\mu}_C+\tilde{\alpha}_C) G^{*3} &+ \left[ \tilde{k}_C\tilde{\mu}_C +  (\tilde{k}_G-1) (\tilde{\mu}_C+\tilde{\alpha}_C) \right] G^{*2} + \\  &+ \left( \tilde{k}_C\tilde{k}_G \tilde{\mu}_C +\tilde{S} - \tilde{k}_C\tilde{\mu}_C - \tilde{k}_G\tilde{\mu}_C  - \tilde{\alpha}_C \tilde{k}_G \right) G^* +
\tilde{S}\tilde{k}_C- \tilde{k}_G \tilde{k}_C \tilde{\mu}_C \Big \}= 0 \, .
\end{aligned}
\end{equation}
One of the solutions is trivially \( G^* = 0 \), while the remaining ones are given by the roots of the cubic polynomial in braces, with at least one root guaranteed to be real. Introducing the following auxiliary quantities
\begin{equation}
\begin{aligned}
p :=& -\dfrac{\left[ \tilde{k}_C\tilde{\mu}_C+\left(\tilde{k}_G-1 \right)\left( \tilde{\alpha}_C+\tilde{\mu}_C\right)\right]^2+3\left(\tilde{\alpha}_C+\tilde{\mu}_C\right)\left[\tilde{\alpha}_C\tilde{k}_G+\left( \tilde{k}_C+\tilde{k}_G-\tilde{k}_C\tilde{k}_G\right)\tilde{\mu}_C-\tilde{S}\right]}{3\left( \tilde{\alpha}_C+\tilde{\mu}_C\right)^2} \, ,\\
q :=& \dfrac{1}{27\left(\tilde{\alpha}_C+\tilde{\mu}_C \right)^3}  
\left\{ 2\left[ \tilde{k}_C\tilde{\mu}_C+\left(\tilde{k}_G-1 \right)\left( \tilde{\alpha}_C+\tilde{\mu}_C\right)\right]^3+27 \tilde{k}_C\left( \tilde{\alpha}_C+\tilde{\mu}_C\right)^2\left( \tilde{S}-\tilde{\mu}_C\tilde{k}_G\right) \right. \notag \\
& \left. -9\left(\tilde{\alpha}_{C}+\tilde{\mu}_C\right)\left[ \tilde{k}_C\tilde{\mu}_C+\left(\tilde{k}_G-1 \right)\left( \tilde{\alpha}_C+\tilde{\mu}_C\right)\right]\left[ \tilde{S}-\tilde{k}_G\left(\tilde{\alpha}_C+\tilde{\mu}_C \right)+\tilde{k}_C\tilde{\mu}_C\left(\tilde{k}_G-1 \right)\right]\right\} \, ,
\end{aligned}
\end{equation}
and defining the discriminant as
\begin{equation}
\Delta := -\left( 4p^3+27q^2\right) \, ,
\end{equation}
the nature of the roots of the cubic can be characterised as follows. If \( \Delta > 0 \), the polynomial has three distinct real roots; if \( \Delta < 0 \), there is one real solution and two complex conjugate roots; and if \( \Delta = 0 \), at least two roots coincide.

It is important to note that only the positive real roots are biologically admissible in this context. Thus, for a given set of parameters, one can evaluate \( p \), \( q \), and \( \Delta \) to determine the number and nature of feasible equilibria. Due to the algebraic complexity of the expressions, the general solution obtained via Cardano's formula is not reported here.

To analyse the system equilibria and their stability, we use the parameters reported in Table~\ref{tab:parameters} to derive the corresponding dimensionless quantities. These values are considered biologically realistic and are drawn from the literature, as previously discussed in Section~\ref{Parameter_List}. The total production rate of cytotoxic T-lymphocytes, \( \tilde{S} \), is left as a free parameter in order to investigate the behaviour of the system under varying therapy intensity. 

Under biologically plausible conditions (i.e., \( G^* \geq 0 \)), the analysis reveals that for values of CTL production rate in a given range, specifically \( \tilde{S}_{cr} < \tilde{S} < \tilde{S}^* \), the system admits three biologically feasible equilibria:
\begin{itemize}
    \item a \textit{healthy equilibrium}, given by \( (G^*, C^*) = (0, \tilde{S}/\tilde{\mu}_C) \);
    \item a \textit{tumour-dominated equilibrium}, associated with a high tumour burden;
    \item a \textit{low-level tumour equilibrium}, characterised by a smaller but non-zero tumour cell population
\end{itemize}


If \( \tilde{S} \leq \tilde{S}_{cr} \), the only biologically feasible equilibria for \( G^* \) are the healthy equilibrium and the tumour-dominated one. As \( \tilde{S} \) increases, corresponding for instance to higher levels of therapy in the model, the low-level tumour equilibrium emerges and, with further increases in therapy, it approaches the tumour-dominated equilibrium until the two coalesce in a saddle-node bifurcation at \( \tilde{S} = \tilde{S}^* \). For values above the turning point $\tilde{S}^*$, i.e., \( \tilde{S} > \tilde{S}^* \), both equilibria disappear, and only the healthy equilibrium persists.

\begin{figure}[ht]
\centering
\includegraphics[width=0.7\linewidth]{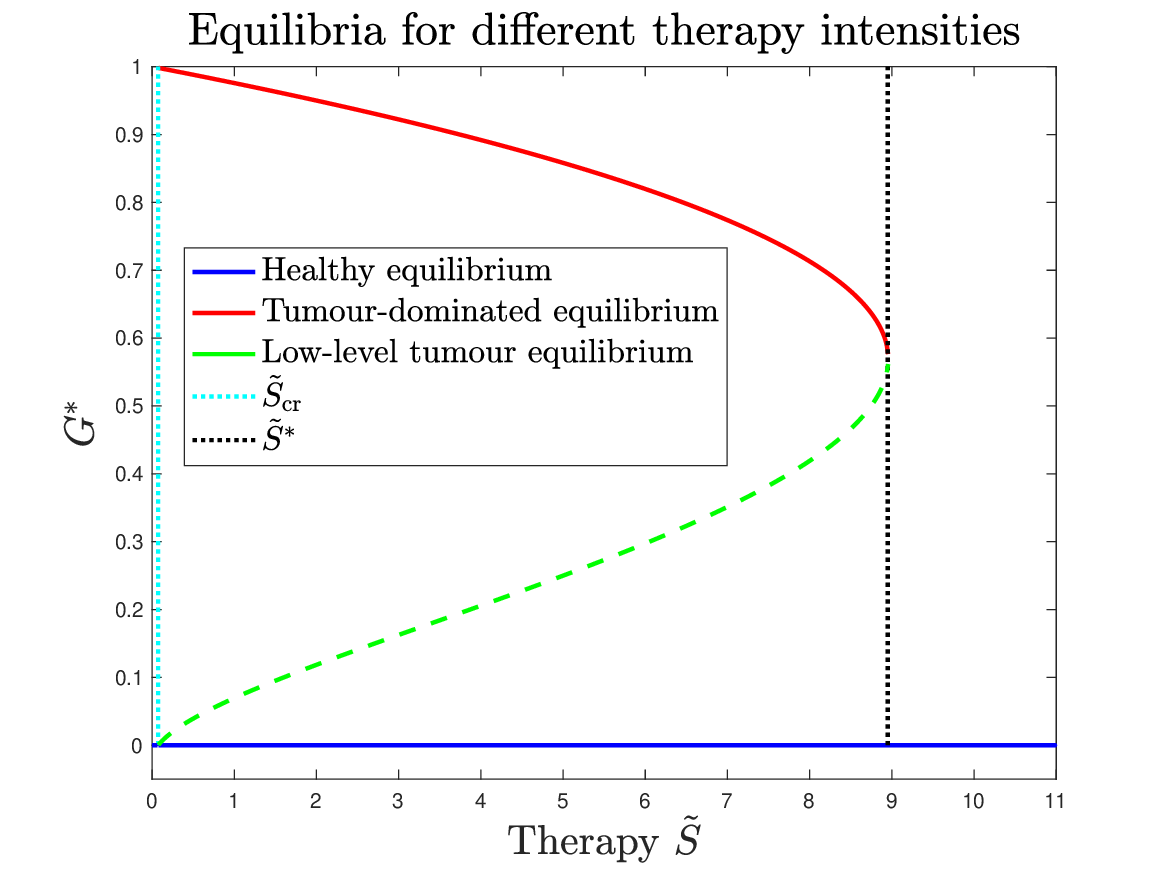}
\caption{Bifurcation diagram highlighting the stability of the identified equilibria, with the model parameters set as in Table~\ref{tab:parameters}. Stable equilibria are depicted with a solid line, whereas unstable equilibria are indicated with a dashed line.}
\label{fig:bifurcation_diagram}
\end{figure}

Figure~\ref{fig:bifurcation_diagram} illustrates the equilibrium values of \( G^* \) as a function of the therapy rate \( \tilde{S} \), with the model parameters set as in Table~\ref{tab:parameters}.
It is important to remember that the values reported in Table~\ref{tab:parameters} are dimensional, whereas the analysis presented here is conducted in a dimensionless framework.

We now proceed to analyse the stability of the equilibria. The Jacobian matrix of the system, evaluated at an equilibrium point \( (G^*, C^*) \), is given by:
\begin{equation}
\label{jacobian}
\left.J\right|_{(G^* \, , \, C^*)}=\left[\begin{array}{cc}
1-2 G^*-C^*\dfrac{\tilde{k}_G}{\left(G^*+\tilde{k}_G\right)^2} & -\dfrac{G^*}{G^*+\tilde{k}_G} \\
-\tilde{\alpha}_C C^*\dfrac{\tilde{k}_C}{\left(G^*+\tilde{k}_C\right)^2}  & -\tilde{\mu}_C-\tilde{\alpha}_C \dfrac{G^*}{G^*+\tilde{k}_C}
\end{array}\right] \, .
\end{equation}
At the healthy equilibrium $(G^* \, , \, C^*)=(0 \, , \, \tilde{S}/\tilde{\mu}_C)$, the Jacobian \eqref{jacobian} reduces to:
\begin{equation}
\label{jacobian_healthy}
\left.J\right|_{(0 \, , \, \tilde{S}/\tilde{\mu}_C)}=\left[\begin{array}{cc}
1-\dfrac{\tilde{S}}{\tilde{\mu}_C \tilde{k}_G}  & 0 \\
-\dfrac{\tilde{\alpha}_C \tilde{S}}{\tilde{\mu}_C \tilde{k}_C}  & -\tilde{\mu}_C
\end{array}\right] \, .
\end{equation}
Given that one eigenvalue is always negative (specifically, equal to \( -\tilde{\mu}_C \)), the healthy equilibrium is stable if and only if the total CTL production rate \( \tilde{S} \) exceeds a critical value:
\begin{equation}
\tilde{S} > \tilde{S}_{\rm cr} := \tilde{k}_G \tilde{\mu}_C \, .
\end{equation}
The value of \( \tilde{S}_{cr} \) represents the lower threshold of the CTL production rate at which a positive, low-level tumour equilibrium emerges. It denotes a subcritical bifurcation point and can be easily derived as the critical value of \( \tilde{S} \) for which \( G^* = 0 \) is a solution of the associated cubic polynomial in eq. \eqref{poly_3}.

The dimensional expression of the critical threshold is \(S_{\rm cr} = k_G \mu_C r \alpha_G^{-1}\), which is found to be lower than biologically admissible values of \( b_C \). This implies that the healthy equilibrium remains stable even in the absence of administered therapy, as the endogenous production term \( b_C \) is sufficient to sustain it.
However, if the immune system is severely compromised (\( b_C < S_{\rm cr} \)), the model predicts that even a small perturbation from \( G^* = 0 \) can trigger tumour onset, underscoring the pivotal role of immune system robustness in preventing tumour progression.

The analysis of the remaining equilibrium branches is not straightforward to address analytically, and a numerical approach was therefore adopted. The system of equations \eqref{C_star}–\eqref{poly_3} was solved numerically, and the resulting non trivial equilibrium values were substituted into the Jacobian matrix to assess local stability via its eigenvalues.
Within the explored range of therapy rates, i.e. \( \tilde{S}_{\rm cr} < \tilde{S} < \tilde{S}^* \), the Jacobian typically exhibits eigenvalues with non-zero real parts, allowing the application of the Hartman–Grobman theorem to determine stability from the sign of the real parts. The analysis shows that the tumour-dominated equilibrium is stable throughout this interval, whereas the low-level tumour equilibrium remains unstable for all values of \( \tilde{S} \in ( \tilde{S}_{\rm cr}, \tilde{S}^* ) \). We also recall that, in this range, the healthy equilibrium is stable as well, implying the existence of a multistable regime.
Figure~\ref{fig:bifurcation_diagram} shows the bifurcation diagram, generated using \textsc{MATLAB}, illustrating the stability properties of the identified equilibria. 

\begin{figure}[ht]
\centering
\includegraphics[width=0.49\linewidth]{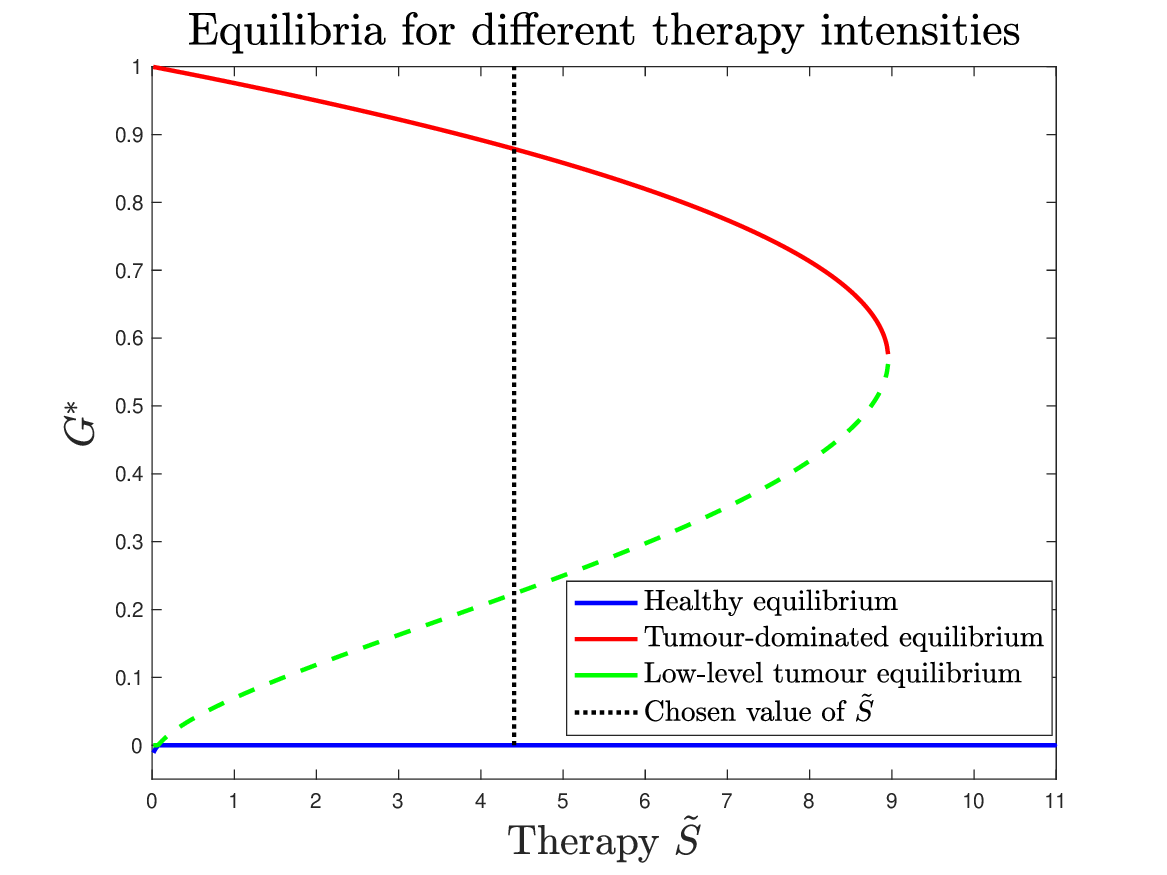} \
\includegraphics[width=0.49\linewidth]{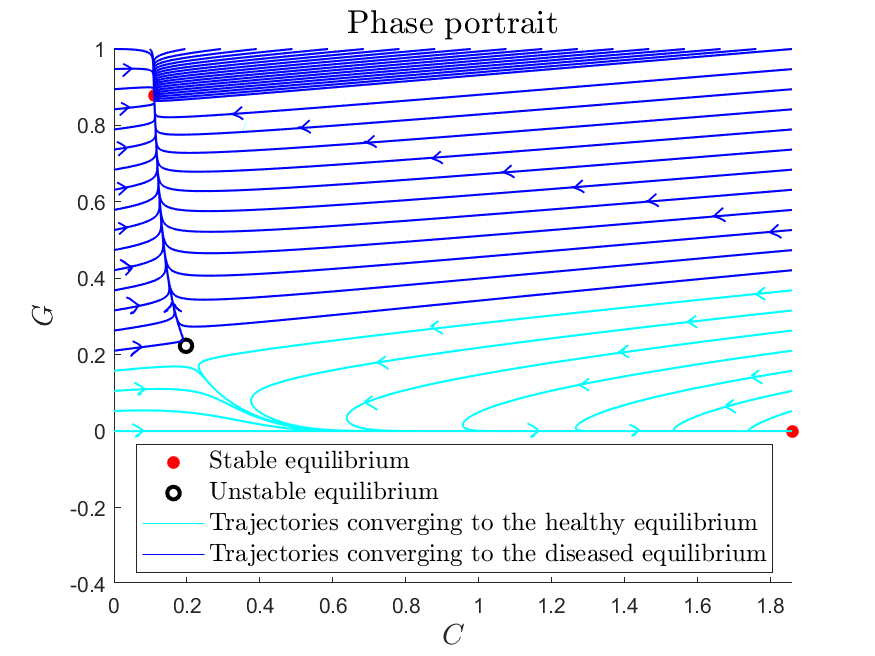}
\caption{Phase portrait for a value of \(\tilde{S}\) where three equilibria exist.}
\label{fig:phase_portrait_1}
\end{figure}

\begin{figure}[ht]
\centering
\includegraphics[width=0.49\linewidth]{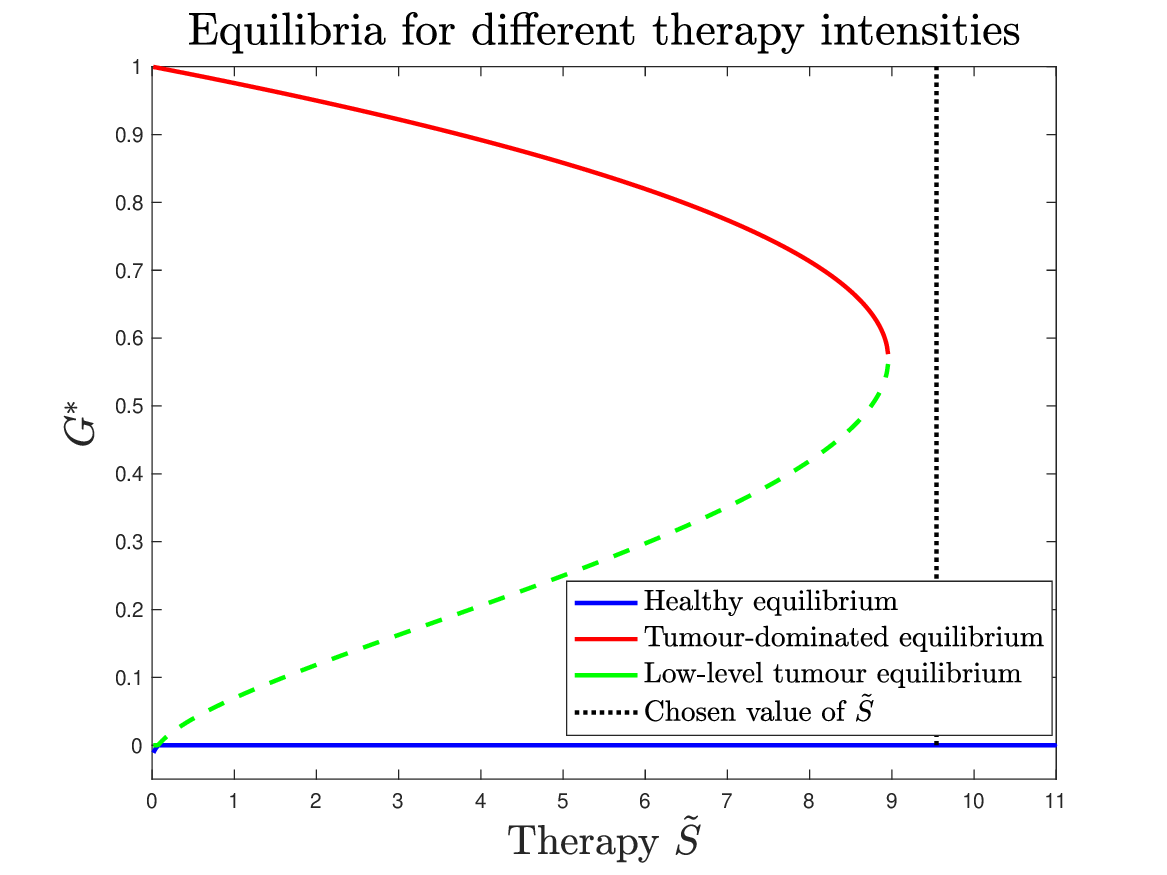} \
\includegraphics[width=0.49\linewidth]{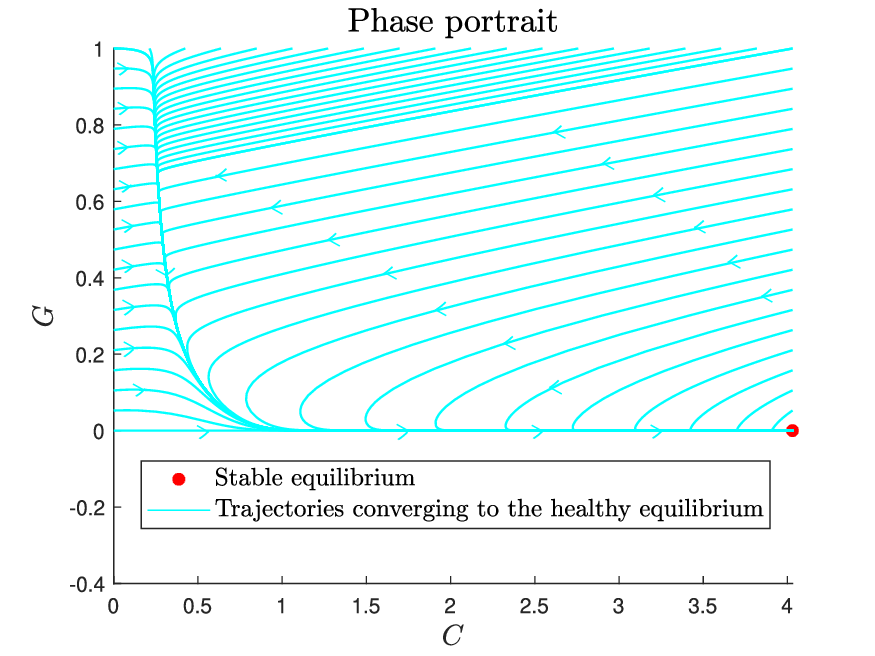}
\caption{Phase portrait for a value of \(\tilde{S}\) where only one equilibrium exists.}
\label{fig:phase_portrait_2}
\end{figure}

Given the complexity introduced by the system’s non-linearities, we investigate numerically the asymptotic behaviour of the system's solutions across the region \(\mathcal{D}\) of the phase space. The software MATLAB is employed to generate phase portraits for various values of the parameter \(\tilde{S}\). The system described by Eq.~\eqref{ODE_adim} is solved numerically over the time interval \([0, 10000]\). Figures \ref{fig:phase_portrait_1} and \ref{fig:phase_portrait_2} illustrate the resulting trajectories for each selected value of \(\tilde{S}\), with multiple initial conditions considered. These phase portraits provide insight into the long-term dynamics of the system and the influence of \(\tilde{S}\) on the stability and behaviour of the trajectories.
To provide a more expressive portrait, the initial conditions are chosen to be evenly spaced along the perimeter of a rectangle in the $G-C$ phase plane, corresponding to the set $[0,1] \times [0,\tilde{S}/\tilde{\mu}_C]$. As previously observed, for values of $\tilde{S}$ between the subcritical bifurcation point and the turning point, i.e. \( \tilde{S}_{\rm cr} < \tilde{S} < \tilde{S}^* \), there are three equilibria. 

As shown in Figure \ref{fig:phase_portrait_1}, the two stable equilibrium points, namely the healthy equilibrium and the tumour-dominated equilibrium, are both attractive, in the sense that almost every initial condition in the phase space generates a trajectory that asymptotically converges to one of them.
In addition, an unstable equilibrium exists along the critical line, acting as a saddle node that delineates the two basins of attraction. On the other hand, for therapy values exceeding the turning point, i.e.,  \(  \tilde{S} > \tilde{S}^* \), there is a single equilibrium point at $(G^*\, , \, C^*) = (0 \, , \, \tilde{S}/\tilde{\mu}_C)$ (see Figure \ref{fig:phase_portrait_2}). Numerical simulations confirm that this point is globally asymptotically stable and acts as an attractor throughout the considered region of the phase space.

\paragraph{Bifurcation analysis with respect to key parameters}

The system’s behaviour is analysed with respect to variations in the key parameters \( \tilde{\alpha}_C \), \( \tilde{k}_G \), \( \tilde{k}_C \), and \( \tilde{\mu}_C \), identified in the previous stability analysis. For each case, the steady states are computed by numerically solving the algebraic system obtained by setting the time derivatives to zero. The local stability of the resulting equilibria is assessed through the eigenvalues of the Jacobian matrix, which are likewise computed numerically.
For each parameter, the bifurcation structure is explored under two different therapy intensities, corresponding to an intermediate and a high therapy level. The two selected values for \( \tilde{S} \) fall within the distinct regimes identified in the previous bifurcation analysis, using the baseline parameters from Table~\ref{tab:parameters}. Specifically, the intermediate value lies within the interval \( [\tilde{S}_{\rm cr},\, \tilde{S}^*] \) previously defined, where multiple biologically feasible equilibria were observed, while the high value corresponds to the regime in which only the healthy equilibrium persists, with the baseline parameters.
This motivates the analysis of how variations in the other parameters may alter the observed behaviour.

\begin{figure}[ht]
\centering
\includegraphics[width=0.49\linewidth]{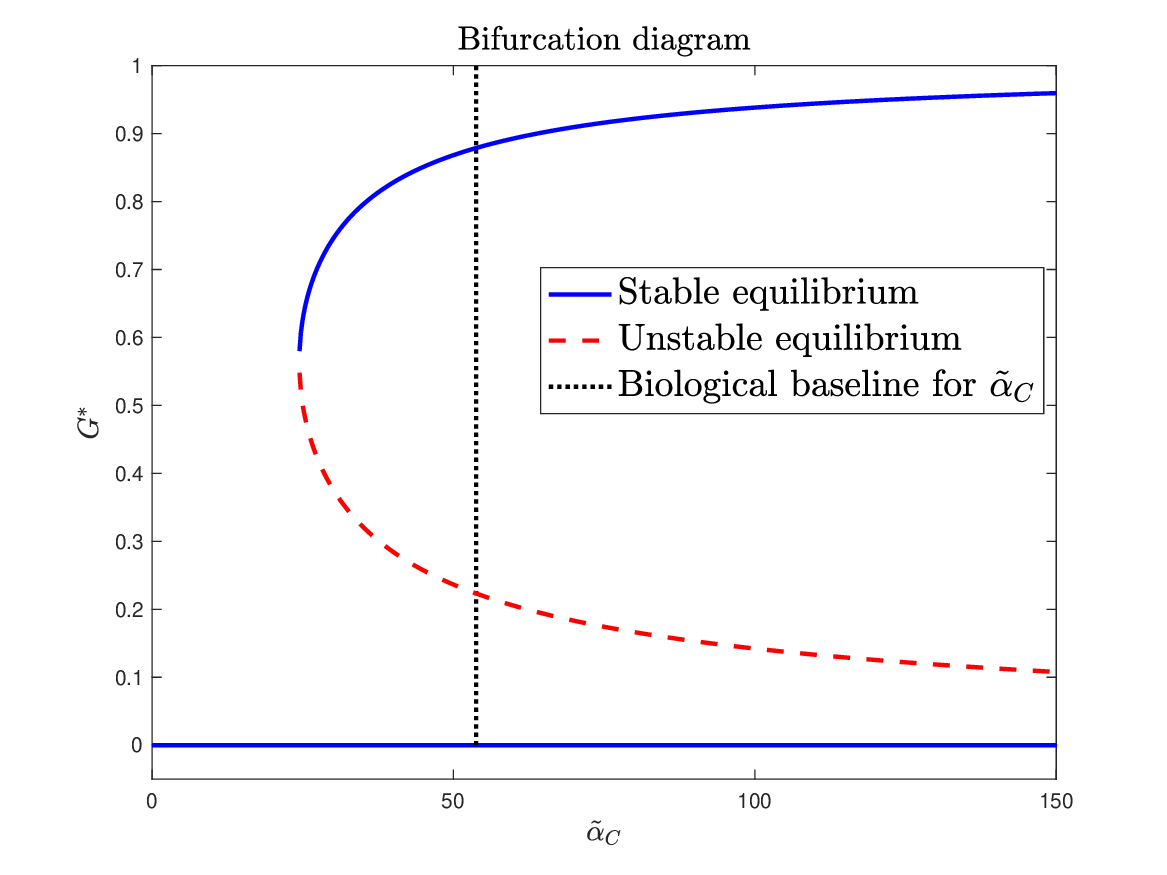} \
\includegraphics[width=0.49\linewidth]{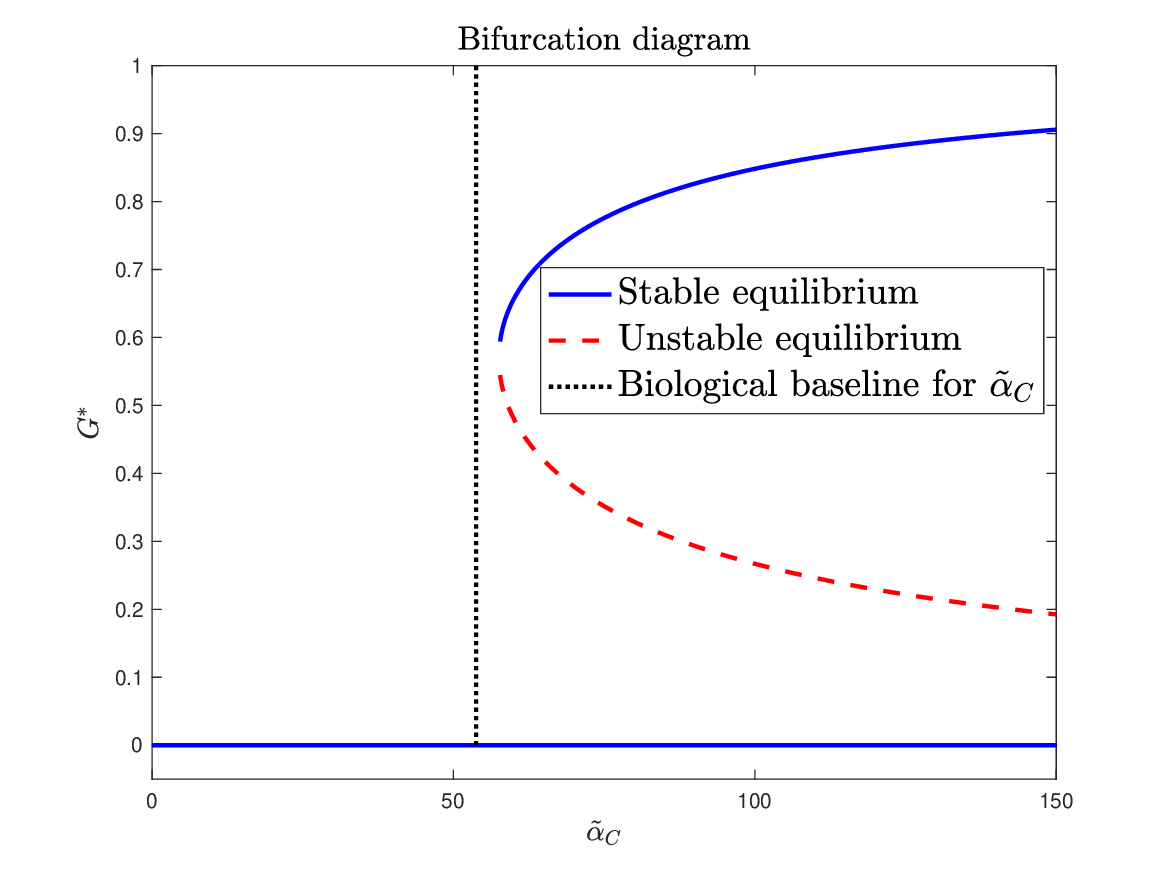}
\caption{Bifurcations diagram of \( \tilde{\alpha}_C \) for the two therapy intensities.}
\label{fig:alpha_C}
\end{figure}

Figure~\ref{fig:alpha_C} presents the bifurcation diagrams with respect to the lymphocyte elimination parameter \( \tilde{\alpha}_C \). 
It follows from \eqref{jacobian_healthy} that the stability of the healthy equilibrium is independent of \( \tilde{\alpha}_C \). As a result, no bifurcations can occur at \( G^* = 0 \) for any value of \( \tilde{\alpha}_C \). For what concerns the other equilibria, for intermediate therapy levels an increase in \( \tilde{\alpha}_C \) causes the tumour-dominated equilibrium to asymptotically approach 1, while the low-level equilibrium tends towards 0. In contrast, as \( \tilde{\alpha}_C \) decreases, these two equilibria collide in a saddle-node bifurcation, beyond which only the healthy equilibrium persists as a biologically feasible steady state. For high therapy levels, the bifurcation diagram retains the overall structure observed in the intermediate therapy case, with the key difference that the saddle-node bifurcation now occurs at values of \( \tilde{\alpha}_C \) exceeding the baseline parameter used in the earlier analysis of \( \tilde{S} \)-variation. As a result, if the lymphocyte elimination rate induced by tumour cells falls below a critical threshold, the system converges to \( G^* = 0 \) as the only stable equilibrium, as previously observed.

\begin{figure}[ht]
\centering
\includegraphics[width=0.49\linewidth]{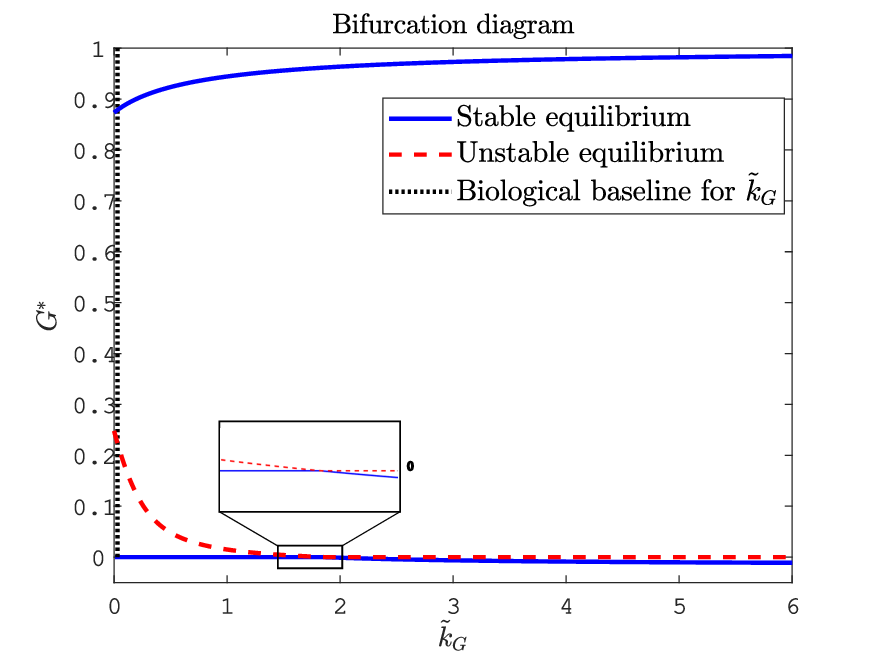} \
\includegraphics[width=0.49\linewidth]{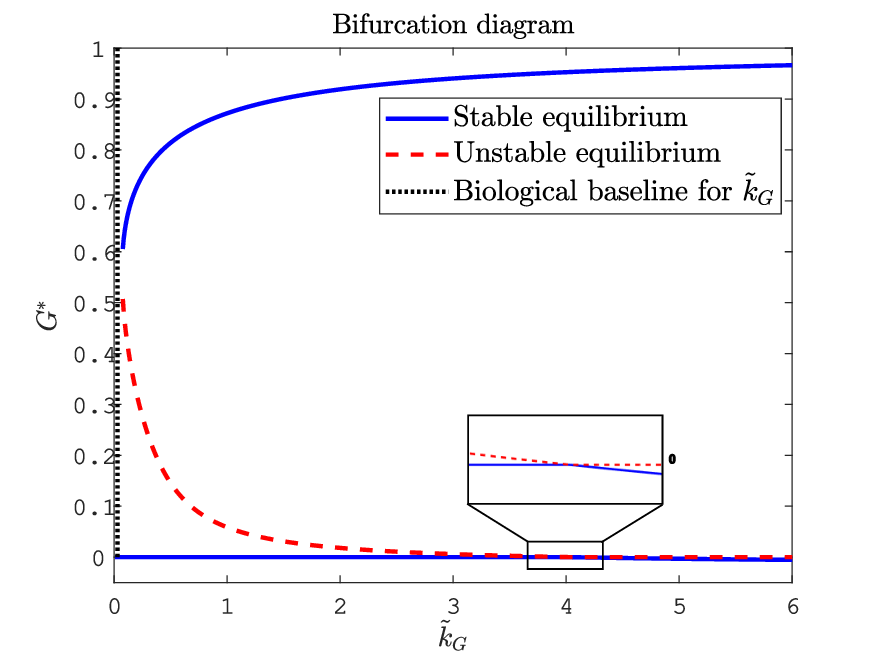}
\caption{Bifurcations diagram of \( \tilde{k}_G \) for the two therapy intensities.}
\label{fig:k_G}
\end{figure}

Figure~\ref{fig:k_G} displays the bifurcation diagrams with respect to the Michaelis–Menten half-saturation constant for tumour cell elimination, \( \tilde{k}_G \). For intermediate therapy levels, increasing \( \tilde{k}_G \) leads to a transcritical bifurcation along the healthy and low-level tumour equilibria branches. This bifurcation causes the low-level tumour equilibrium to disappear (being negative values non-physical) and destabilises the healthy state, while the tumour-dominated equilibrium remains stable and asymptotically approaches 1. At higher therapy levels, the system initially exhibits only the healthy equilibrium at the baseline parameter value. However, as \( \tilde{k}_G \) increases, a pair of equilibria emerge via a saddle-node bifurcation: a stable tumour-dominated state and an unstable low-level state. As \( \tilde{k}_G \to +\infty \), the diseased equilibrium tends to 1, while the unstable branch eventually collides with the healthy equilibrium in a transcritical bifurcation, rendering \( G^* = 0 \) unstable.
Overall, this analysis shows that for large values of \( \tilde{k}_G \), the only biologically meaningful and stable equilibrium is the tumour-dominated one. This behaviour arises because, in such a regime, the system operates far from saturation, resulting in a lymphocyte-induced tumour cell elimination rate that is insufficient to prevent tumour persistence.

\begin{figure}[ht]
\centering
\includegraphics[width=0.49\linewidth]{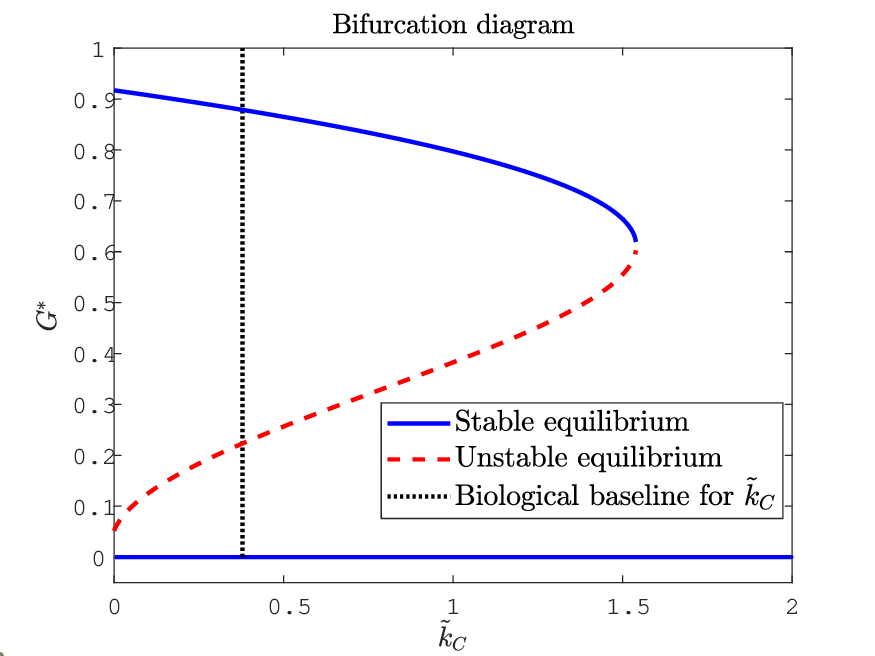} \
\includegraphics[width=0.49\linewidth]{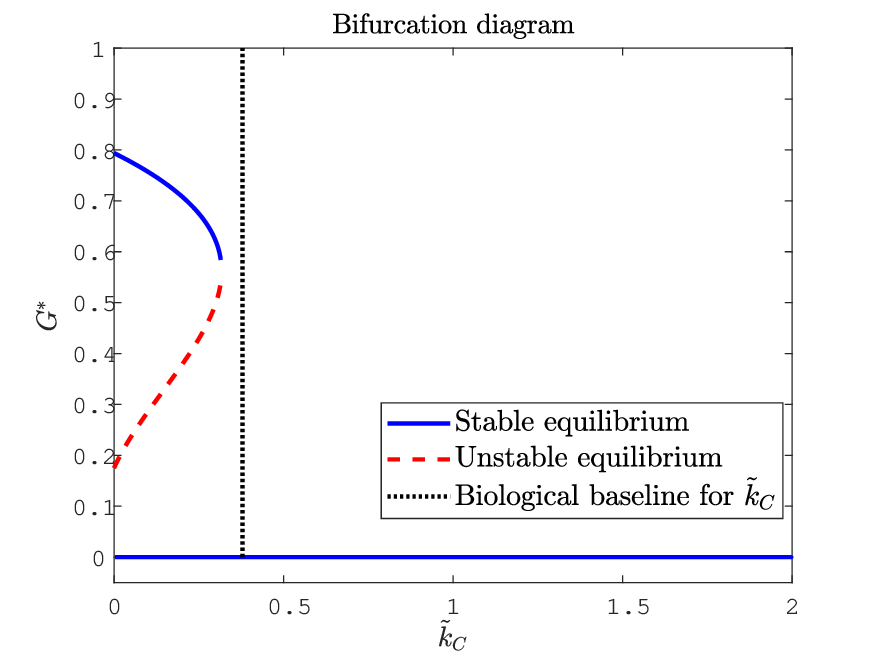}
\caption{Bifurcations diagram of \( \tilde{k}_C \) for the two therapy intensities.}
\label{fig:k_C}
\end{figure}

Then, from the Jacobian expression \eqref{jacobian_healthy}, it follows that the Michaelis–Menten half-saturation constant for lymphocyte-induced tumour cell clearance, \( \tilde{k}_C \), does not affect the sign of the eigenvalues at the healthy equilibrium. Consequently, no bifurcations occur at \( G^* = 0 \) as a function of \( \tilde{k}_C \), and the healthy equilibrium remains stable for all values of this parameter. For low values of \( \tilde{k}_C \), the tumour-dominated equilibrium is also stable, while the low-level tumour equilibrium is unstable. As \( \tilde{k}_C \) increases, these two equilibria coalesce and disappear via a saddle-node bifurcation, leaving the healthy equilibrium as the sole biologically feasible steady state. The therapy intensity \( \tilde{S} \) influences the range of \( \tilde{k}_C \) for which the tumour-dominated regime is observed and the saddle-node bifurcation occurs, but does not alter the overall bifurcation structure. The corresponding bifurcation diagrams are presented in Figure~\ref{fig:k_C}.

\begin{figure}[ht]
\centering
\includegraphics[width=0.49\linewidth]{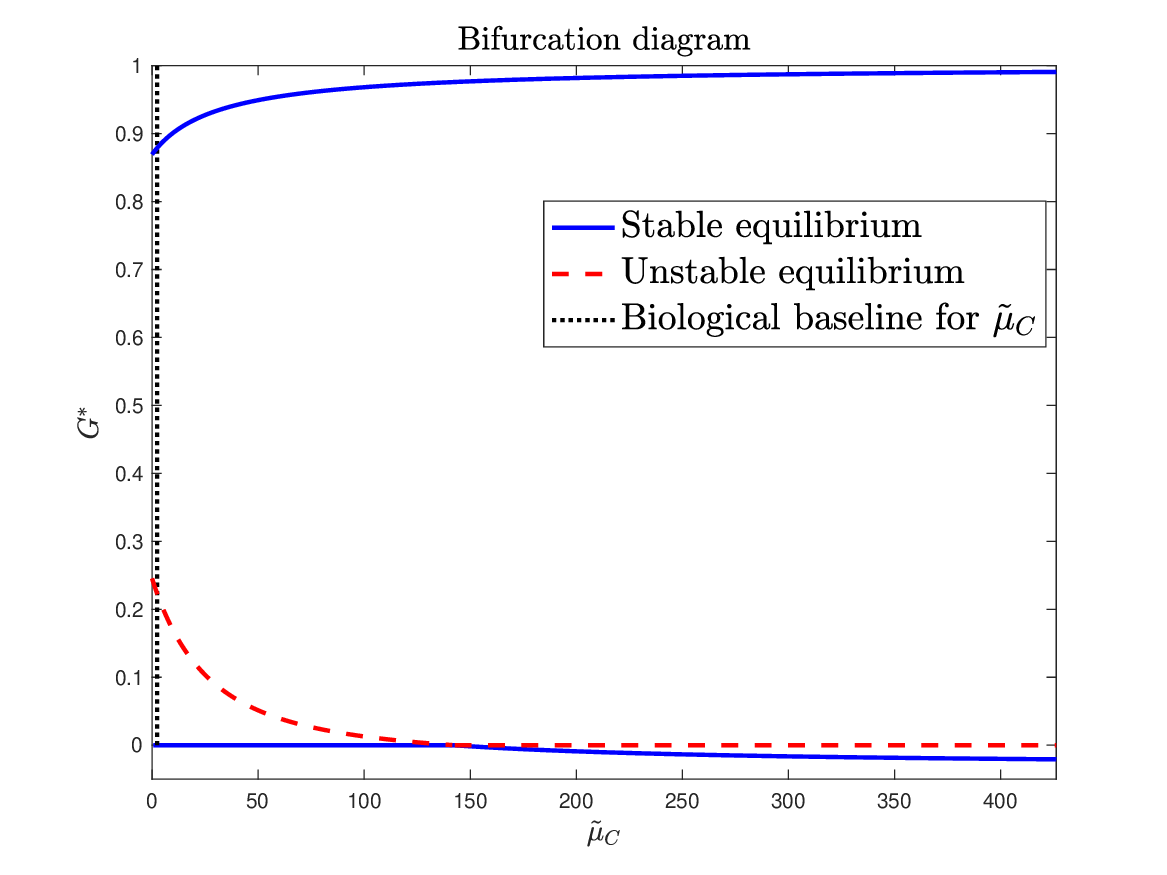} \
\includegraphics[width=0.49\linewidth]{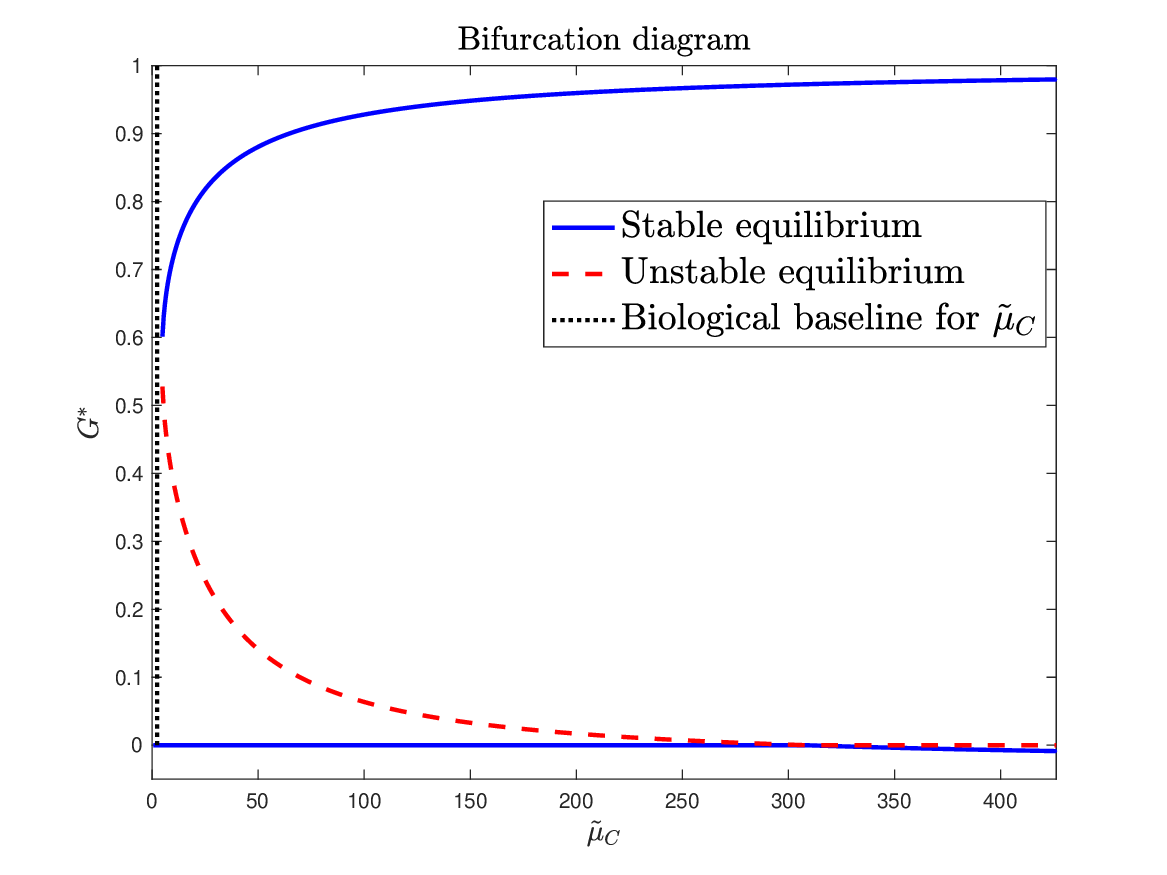}
\caption{Bifurcations diagram of \( \tilde{\mu}_C \) for \( \tilde{S} = 60 \tilde{S}_{\rm cr} \) and \( \tilde{S} = 130 \tilde{S}_{\rm cr} \).}
\label{fig:mu_C}
\end{figure}

Finally, the bifurcation diagrams for the lymphocyte decay rate parameter, \( \tilde{\mu}_C \), are shown in Figure~\ref{fig:mu_C}. In the case of the intermediate value of therapy, for \( \tilde{\mu}_C \) above a threshold (lower than the baseline given in Table \ref{tab:parameters}), the system admits three biologically feasible equilibria: a stable healthy equilibrium, an unstable low-level tumour equilibrium, and a stable tumour-dominated equilibrium. As \( \tilde{\mu}_C \) increases, a transcritical bifurcation occurs at the healthy branch, rendering the equilibrium at \( G^* = 0 \) unstable. The resulting stable equilibrium becomes negative and thus biologically unfeasible. In the case of the higher therapy value, the system exhibits only the healthy equilibrium for values of \( \tilde{\mu}_C \) less than or equal to the baseline. As \( \tilde{\mu}_C \) increases, a saddle-node bifurcation generates a stable tumour-dominated equilibrium and an unstable low-level equilibrium. For sufficiently large values of \( \tilde{\mu}_C \), a transcritical bifurcation at \( G^* = 0 \) occurs, destabilising the healthy equilibrium, while the low-level equilibrium becomes negative and loses physical relevance.

\paragraph{Solution of the system of ODEs.}

The results of the model simulations are now presented for selected test cases, obtained by numerically solving system~\eqref{ODE_adim} in MATLAB. The parameter values used are those reported in Table~\ref{tab:parameters}, and various therapy levels and tumour initial conditions are considered. In all simulations, the initial lymphocyte concentration is set to \( C(0) = \tilde{b}_C \).

\begin{figure}[ht]
\centering
\includegraphics[width=0.49\linewidth]{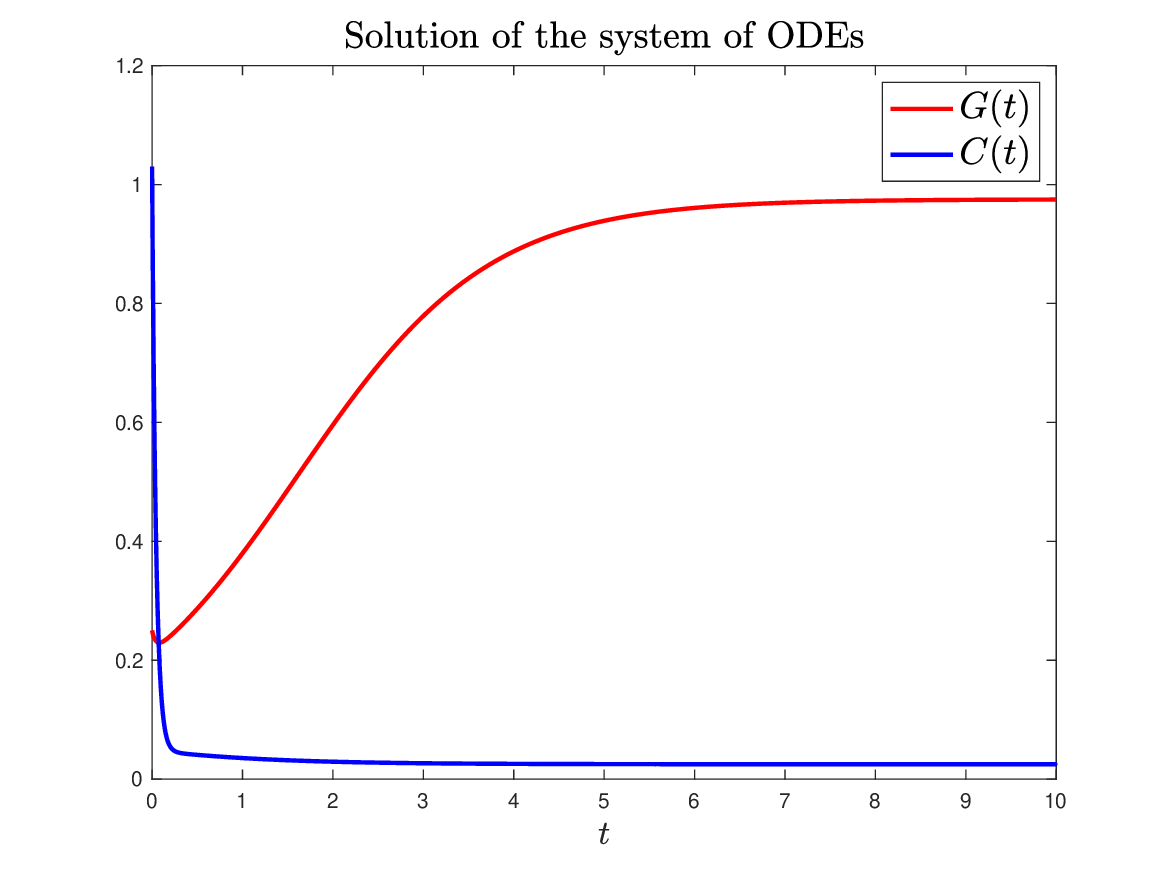}
\includegraphics[width=0.49\linewidth]{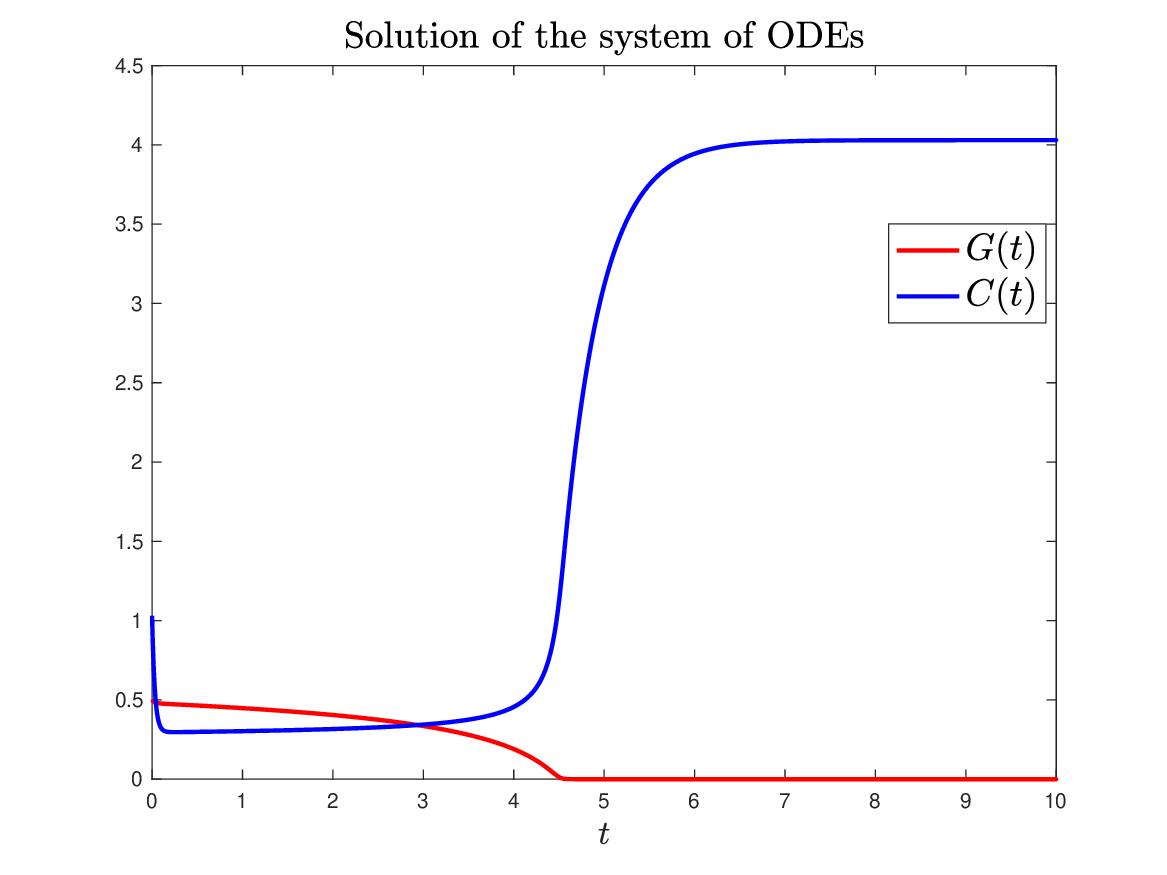}
\caption{The evolution of the system for \(\tilde{S}= \tilde{b}_C \) and an initial tumour condition of \( G(0) = 0.25 \), on the left, and for \( \tilde{S} = 130\tilde{S}_{\rm cr} \) and an initial tumour condition of \( G(0) = 0.5 \), on the right.}
\label{fig:ODE_S=b_C}
\end{figure}

Figure~\ref{fig:ODE_S=b_C}-left shows the evolution of tumour and lymphocyte concentrations in the absence of therapy (\( \tilde{S} = \tilde{b}_C \)), with an initial tumour burden of \( G(0) = 0.25 \). As observed, the immune response alone is insufficient to control tumour progression, which grows close to its carrying capacity.
On the other hand, Figure~~\ref{fig:ODE_S=b_C}-right illustrates the trajectories of \( G \) and \( C \) for therapy intensity higher than $\tilde{S}^{*}$ (e.g. \( \tilde{S} = 130\, \tilde{S}_{\rm cr}\) ), and an initial tumour concentration of \( G(0) = 0.5 \). In this setting, the therapy consistently suppresses tumour growth, driving the system toward tumour eradication.

\begin{figure}[ht]
\centering
\includegraphics[width=0.49\linewidth]{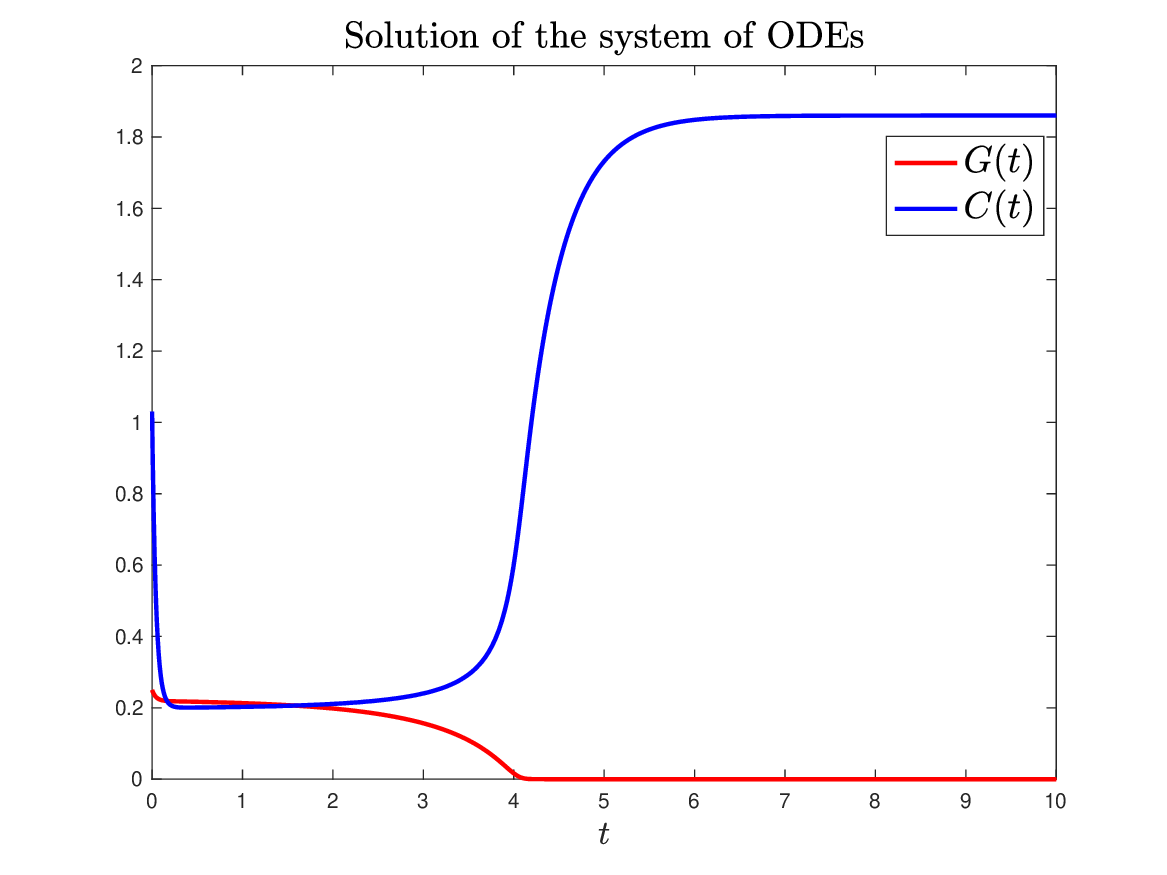} \
\includegraphics[width=0.49\linewidth]{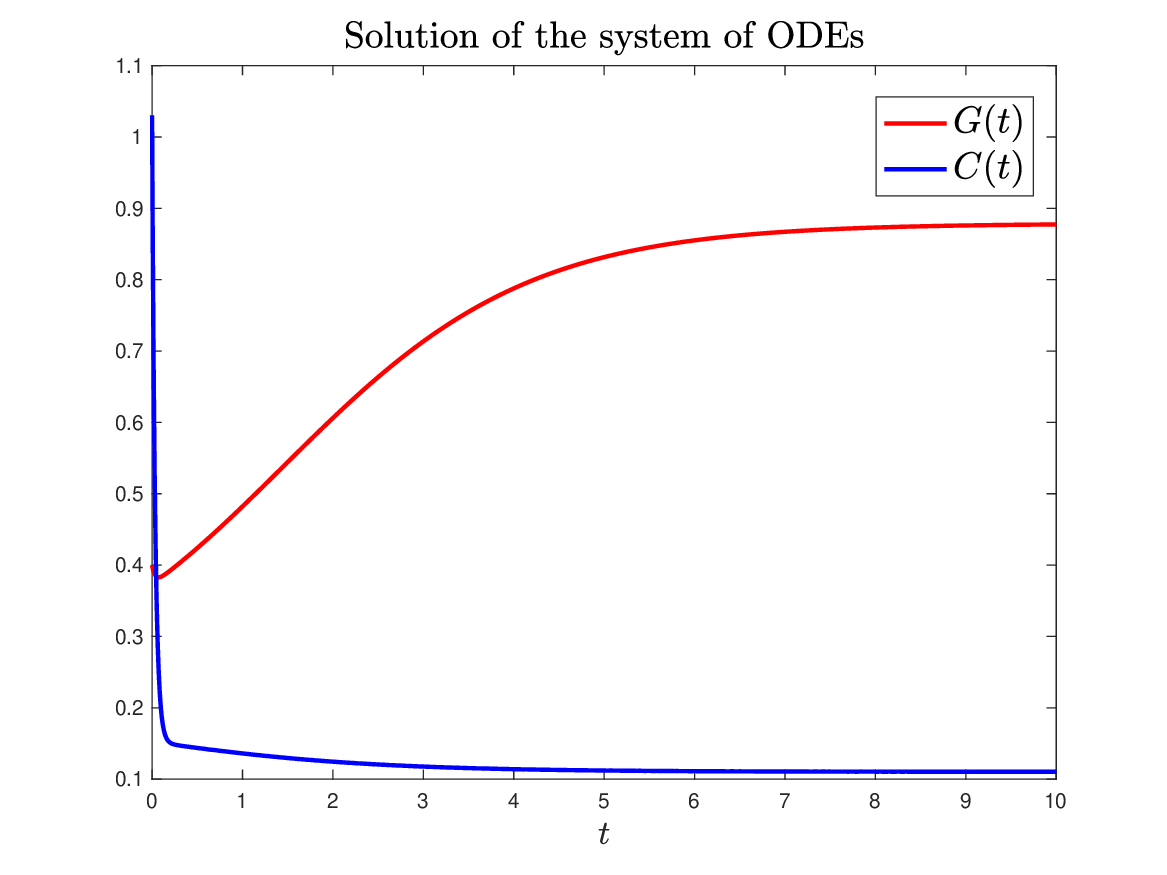}
\caption{The evolution of the system for \( \tilde{S} = 60\tilde{S}_{\rm cr} \) and an initial tumour condition of \( G(0) = 0.25 \) and \( G(0) = 0.4 \).}
\label{fig:ODE_S=60}
\end{figure}

The system dynamics for a therapy intensity in a regime where both tumour eradication and persistence are possible (e.g. \( \tilde{S} = 60\, \tilde{S}_{\rm cr} <  \tilde{S}^{*} \) ), is reported in  Figure~\ref{fig:ODE_S=60}. Two different initial tumour conditions are considered: \( G(0) = 0.25 \), on the left, and \( G(0) = 0.4 \), on the right. In the first case, therapy successfully leads to tumour eradication (see Fig.~\ref{fig:ODE_S=60}-left), while in the second case, the treatment is not sufficient to prevent tumour growth (see Fig.~\ref{fig:ODE_S=60}-right).

These simulation results are fully consistent with the outcomes of the stability and bifurcation analysis presented earlier, confirming the presence of multiple attractors for intermediate therapy levels and the effectiveness of higher therapy rates in ensuring tumour clearance.


\section{Numerical simulations}
\label{sec:numerical_simulations}
In this Section, the model represented by system \eqref{dimentional_PDE_model}, as introduced in Section \ref{sec:mathematical_model}, is solved numerically. Initially, the results of model simulations in a simplified 2D geometry will be presented. The relative importance of different parameters will be assessed through a sensitivity analysis on the simulation outcomes conducted with various parameter sets. Subsequently, the system will be simulated in 3D within the brain geometry, using patient-specific DTI data to identify preferential directions. All simulations were carried out using FEniCS, an open-source computing platform for solving partial differential equations (PDEs) using the finite element method.

\subsection{Two dimensional simulations on a simplified domain} 

We begin by investigating tumour growth within a simplified two-dimensional domain, with a particular focus on the effects of therapy application. All results and plots are presented in terms of rescaled concentrations: the tumour cell concentration is normalised by its carrying capacity, while the concentrations of lymphocytes and TGF-$\beta$ are scaled by the constants \( C_{m} := 10^6 \, \text{cells/mL} \) and \( T_{m} := 10 \, \text{pg/mL} \), respectively.

Finally, in the 2D simulations, diffusion is assumed to be isotropic, with the diffusion tensors set as $\mathbb{D}_G = D_G \mathbb{I}$, $\mathbb{D}_C = D_C \mathbb{I}$, and $\mathbb{D}_T = D_T \mathbb{I}$, with $D_G$, $D_C$, and $D_T$ as in Table \ref{tab:parameters}.

\subsubsection{Numerical simulation setup}
\label{2D_setup}
\paragraph{Initial conditions and therapeutic input.}
The initial tumour cell concentration is a Gaussian centred at the origin of the XY plane: 
\begin{equation}\label{CI_G}
    G(x,y,t=0)= G_0\exp\left(-\dfrac{x^2+y^2}{2\sigma_G^2}\right) \, ,
\end{equation}
with \(\sigma_G=0.124 \text{ cm}\) and \(G_0 = 0.5\), meaning that the initial maximum concentration is half of the carrying capacity. For both lymphocyte and chemoattractant, the initial concentrations are set to zero:
\begin{equation}
C(x,y,t=0)\equiv 0 \, , \quad 
T(x,y,t=0)\equiv 0 \, .
\end{equation}

Lymphocytes are then delivered as part of the immunotherapy treatment. In particular, the use of lymphocyte infusion for the treatment of GBM is currently under investigation in clinical trials, especially in combination with CAR-T cell engineering techniques~\cite{Pinheiro:2023}. Several delivery strategies have been proposed: T-cells can be administered peripherally via intravenous infusion~\cite{Ahmed:2017}, attracted to the brain by chemotactic signals that help them cross the blood–brain barrier (BBB)~\cite{Banerjee:2015}, or infused directly into the brain parenchyma through a catheter~\cite{Brown:2016}.
In this work, the latter strategy is implemented by prescribing a spatially concentrated lymphocyte source, where the infusion term \(S_T(x, y)\) 
is modelled as a time-constant Gaussian profile
\begin{equation}\label{S}
{S}_T(x,y)= S_0 \exp\left(-\dfrac{(x-x_T)^2+(y-y_T)^2}{2\sigma_S^2}\right)  \, ,
\end{equation}
with \(\sigma_S=0.044 \text{ cm}\). Therapy intensities are modulated by adjusting \(S_0\). The centre of the Gaussian represents the catheter tip position for therapy infusion, set at \((x_T,y_T)=(1.1 \text{ cm}, 1.1 \text{ cm})\). 

\begin{figure}[ht]
\centering\includegraphics[width=0.75\linewidth]{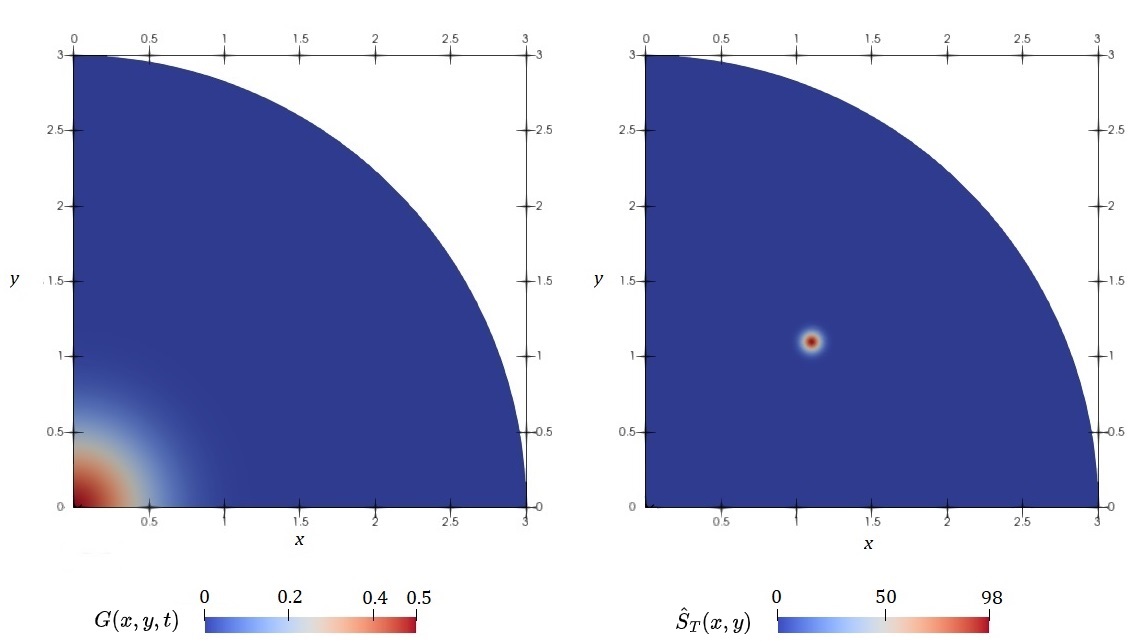}
\caption{Initial conditions for tumour concentration $G(x,y,t=0)$ and therapy source term \({S}_T(x,y)\) for \(S_0=10^2 \ {\rm h}^{-1}\), where therapy values are rescaled with respect to \(C_m\).}
\label{initial_conditions}
\end{figure}

The initial conditions for tumour concentration and for the therapy source term are shown in Fig.~\ref{initial_conditions}.

\paragraph{Mesh creation.}

\begin{figure}[t]
\centering
\includegraphics[width=0.85\linewidth]{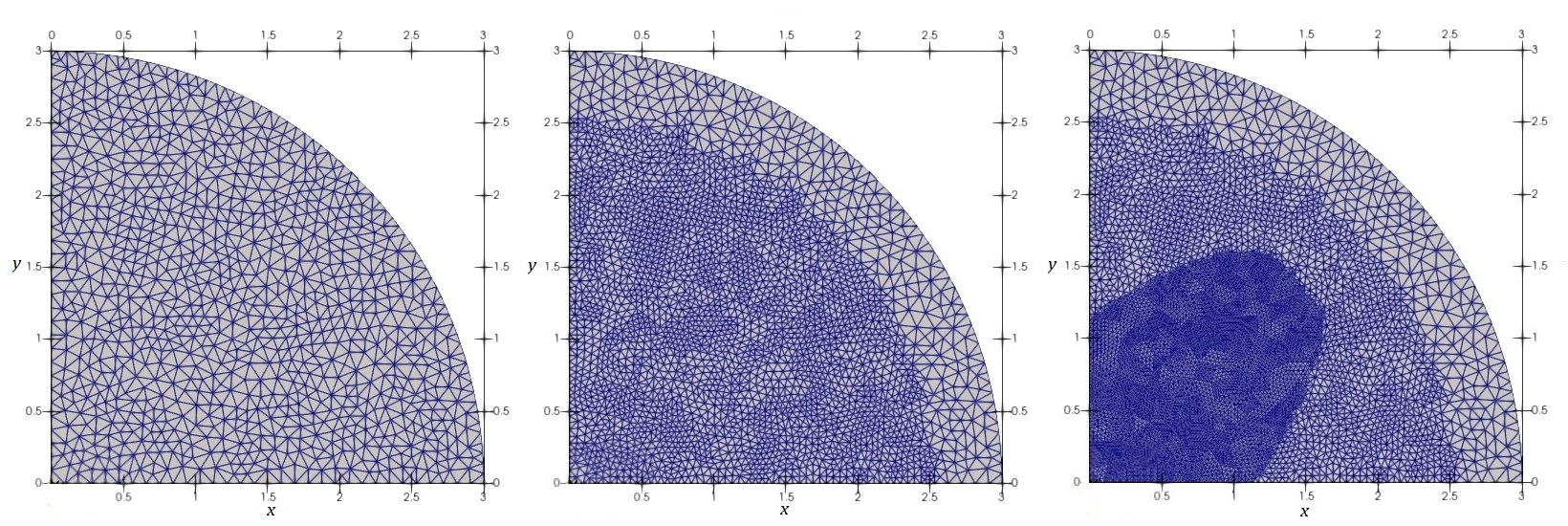}
\caption{Computational mesh generated in FEniCS, shown after the initial automatic generation and after each of the two refinement steps.}
\label{fig:mesh2D}
\end{figure}

Regarding the computational domain, it is assumed that under conditions of free growth, the tumour maintains radial symmetry. To reduce computational costs, the domain is restricted to a circular sector of \(90^\circ\). Simulations are run up to a final time of 2400 hours (corresponding to 100 days, the median survival time for untreated GBM). Given that a typical GBM reaches a diameter of approximately 3 cm at the time of patient death, the radius of the circular sector is set to this value. The computational mesh is automatically generated in FEniCS and subsequently refined through two adaptive refinement steps where needed. The mesh and its refinements are illustrated in Figure~\ref{fig:mesh2D}. 
As a first step, taking into account the expected spatial expansion of tumour cells over the simulation time, mesh refinement is applied within a circular region of radius \( R = 2.5 \) cm.

Moreover, the transport term may introduce numerical instabilities, which are often handled using stabilisation techniques. In this study, we mitigate these effects by employing a sufficiently refined mesh, which helps control spurious oscillations.
Given the radial symmetry of the domain, a significant lymphocyte flux is expected along the segment connecting the tumour centre to the site of therapy administration. The second refinement step therefore targets mesh cells in this region, with a sufficiently large margin to ensure numerical robustness, especially as parameters vary.

\paragraph{Chemotactic coefficient estimation.} Regarding the chemotactic coefficient, whose estimation remains necessary, we initially assume a constant value \( \chi \), derived from a conservative lymphocyte migration speed of \( v = 4 \, \mu\text{m/min} = 0.024 \, \text{cm/h} \) \cite{Letendre:2015}, and a computed maximal TGF-\(\beta\) gradient of \( \overline{\nabla T} \approx 4.9 \, \text{cm}^{-1} \), obtained from a simulation in which the tumour grows in the absence of treatment. This yields an approximate value \( \chi \approx 4.9 \cdot 10^{-3} \, \text{cm}^2/\text{h} \). However, simulations under this assumption showed that lymphocytes tended to concentrate in the tumour centre rather than remaining near its boundary.

To better reflect the expected biological behaviour, we therefore adopted a receptor-saturation-inspired expression for the chemotactic coefficient~\cite{Hillen:2008}:
\begin{equation}
\label{chemotactic_coefficient_2D}
\chi(G, C, T) = \frac{\tilde{\chi}}{(1 + \epsilon T)^2} \, ,
\end{equation}
where \( \tilde{\chi} = 4.9 \cdot 10^{-3} \, \text{cm}^2/\text{h} \) as before, and \( \epsilon \) is a saturation parameter. Based on the estimated TGF-\(\beta\) concentration at the tumour boundary (approximately 0.35) and assuming a 70\% reduction in migration speed due to receptor saturation, we set \( \epsilon = 0.56 \).

\paragraph{Time discretization.}

To ensure numerical stability in the presence of advective terms, the time step was selected to satisfy the Courant–Friedrichs–Lewy (CFL) condition. In the current setup, the maximum expected lymphocyte velocity is estimated as \( u \leq 0.024\, \text{cm/h} \), and the smallest mesh element has a size of \( h_{\text{min}} = 0.019\, \text{cm} \). A conservative time step of \( \Delta t = 0.2\, \text{h} \) was adopted, resulting in a Courant number of approximately \( C_T \sim 0.258 \), well below the stability threshold. This time step was also used in all simulations of the sensitivity analysis, where larger values of the chemotactic coefficient \(\tilde{\chi}\) were explored. Additionally, a runtime check was implemented in the numerical code to ensure that the CFL condition remains satisfied throughout the simulations.

\subsubsection{Numerical Results under Baseline Conditions} \label{chemio_2D}

\begin{figure}
\centering
\includegraphics[width=0.8\linewidth]{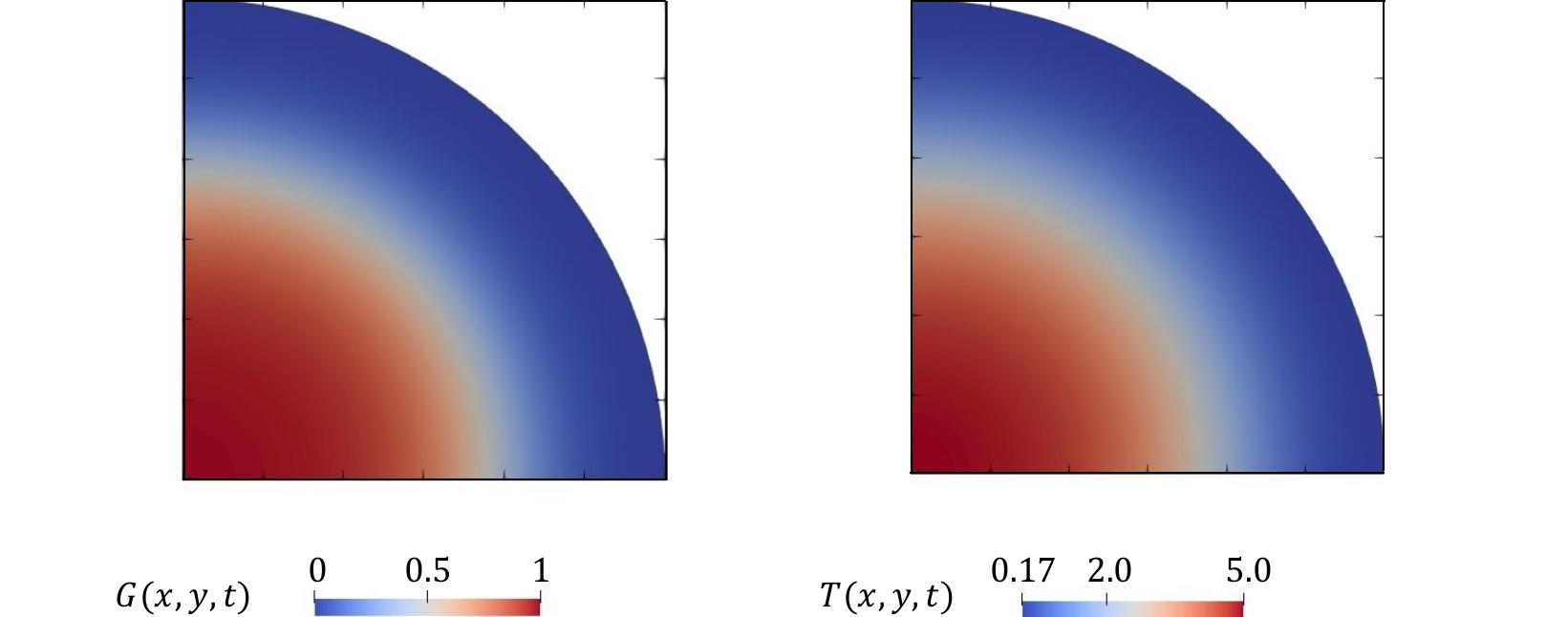}
\caption{Tumour cell concentration \(G(x,y,t)\) and TGF-$\beta$ concentration \(T(x,y,t)\) at $t=2400$ h in the absence of any therapy.
}
\label{fig:2D_G_T_noter}
\end{figure}

\begin{figure}
    \centering
    \includegraphics[width=\linewidth]{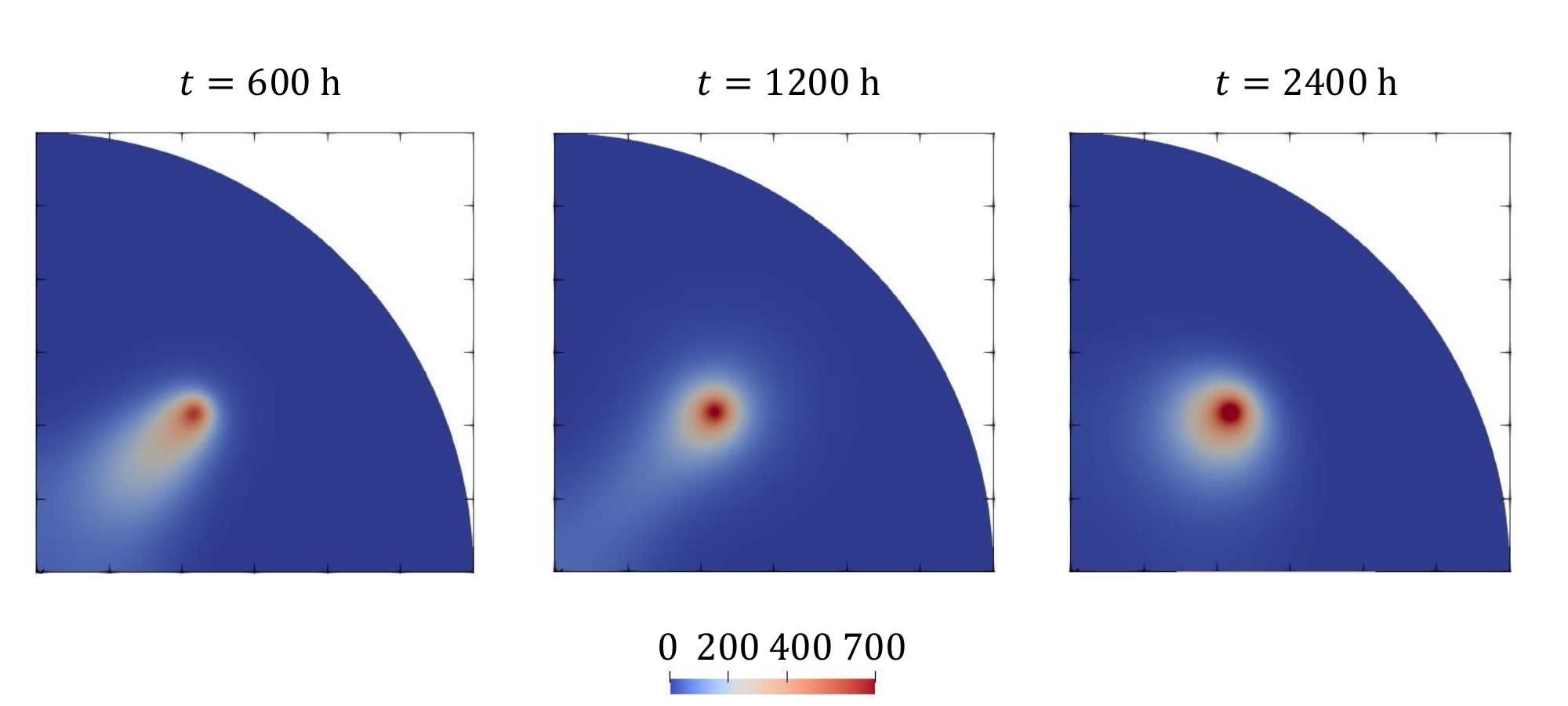}
    \caption{Lymphocyte concentration \( C(x, y, t) \) at various time points, expressed relative to \( C_m \).}
    \label{fig:concentration_C}
\end{figure}

The system \eqref{dimentional_PDE_model} is solved using the parameter values outlined in Section~\ref{Parameter_List}. 
Figure~\ref{fig:2D_G_T_noter} shows the results of tumour cell concentration \(G(x,y,t)\) and TGF-$\beta$ concentration \(T(x,y,t)\), obtained using the variable chemotactic coefficient described in \eqref{chemotactic_coefficient_2D} and setting \( \alpha_C = \alpha_G = 0 \) to better isolate and observe lymphocyte migration.
Figure~\ref{fig:concentration_C} presents the simulation results for the lymphocyte concentration \( C \). As shown, lymphocytes predominantly accumulate in the peripheral regions of the tumour, avoiding excessive infiltration into the hypoxic core. This behaviour is consistent with a more biologically realistic spatial distribution, resulting from receptor saturation effects.

\begin{figure}[!ht]
\centering
\includegraphics[width=0.8\linewidth]{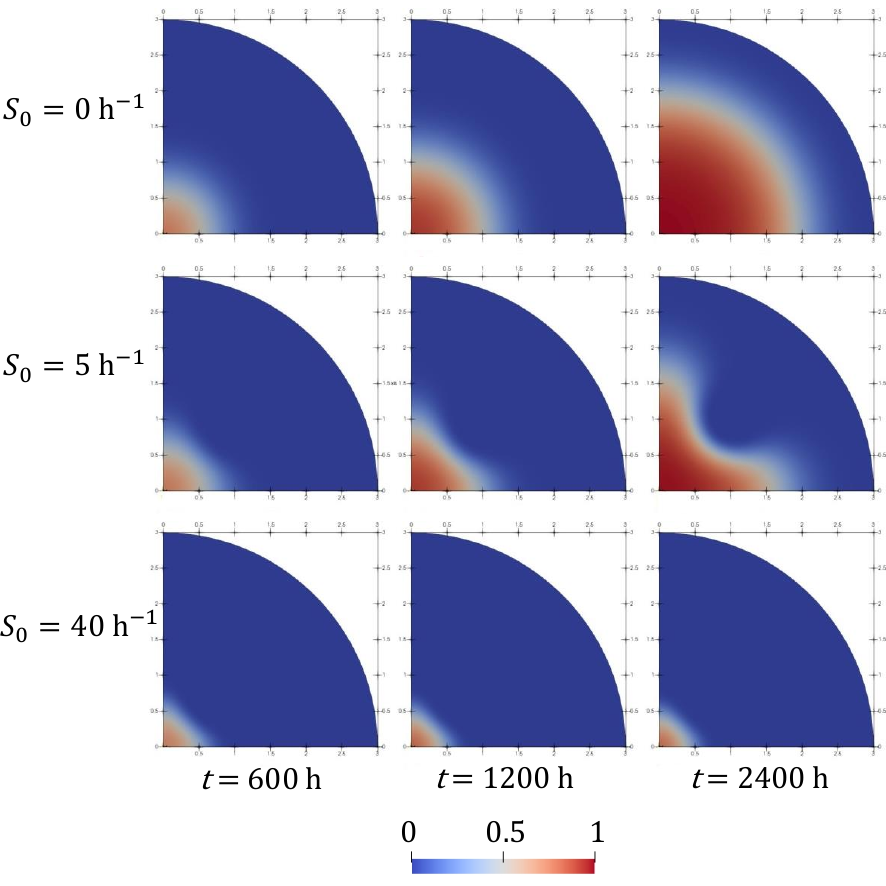}
\caption{Tumour cell concentration \(G(x,y,t)\) at different time instants for rescaled therapy intensities \(S_0 = 0, \, 5, \, 40 \, \, \text{h}^{-1}\), with therapy values expressed relative to \(C_m\).
}
\label{fig:2D_G}
\end{figure}

Finally, the results of the simulation of the full system \eqref{dimentional_PDE_model} with the saturation-dependent chemotactic coefficient are reported in Figure~\ref{fig:2D_G}, in terms of the tumour cell concentration at \( t = 600 \, \mathrm{h} \), \( t = 1200 \, \mathrm{h} \), and \( t = 2400 \, \mathrm{h} \), under varying therapy intensities: \( S_0 = 0 \, \mathrm{h}^{-1} \), \( S_0 = 20 \, \mathrm{h}^{-1} \), and \( S_0 = 40 \, \mathrm{h}^{-1} \). These scenarios highlight the differential effects of therapy on tumour growth. Specifically, the first simulation demonstrates uncontrolled tumour proliferation in the absence of therapy. In the second case, therapy slows tumour expansion but does not achieve adequate containment. Conversely, the highest therapy intensity effectively suppresses tumour growth, resulting in a final tumour size comparable to the initial one.

\subsubsection{Sensitivity analysis of model parameters}
\label{sensitivity_analysis}

After presenting the results obtained in the default parameter setting, we now explore how variations in key model parameters affect the tumour–immune dynamics. Indeed, the proposed model incorporates various parameters obtained from multiple sources. A sensitivity analysis is conducted to determine which parameters significantly influence the model's outcomes and to assess the robustness of the model results under different biological assumptions.

Following the procedure outlined in \cite{Marino:2008}, the \emph{partial rank correlation coefficient} (PRCC) approach is used. The process involves selecting \(M\) parameters of interest, with corresponding vectors \(\boldsymbol{z_j} = \{z_{j1}, z_{j2}, \ldots, z_{jN}\}\), for \(j = 1, \ldots, M\), each containing equispaced values from a specified interval, which are then randomly permuted. These are used to generate a matrix referred to as the \emph{Latin hypercube sample} (LHS) matrix, of dimension \(N \times M\). Each column of the matrix corresponds to one parameter and contains stratified samples from its interval of variation. Each row represents a complete set of parameter values, defining one simulation run in the sensitivity analysis.

The model is solved numerically for each parameter combination defined by a row of the LHS matrix, and an appropriate measure of the numerical output is selected to obtain a real-valued objective function. The results of these simulations form the vector \(\boldsymbol{y} = \{y_1, y_2, \ldots, y_N\}\). For each parameter vector \(\boldsymbol{z_j}\), the Pearson correlation coefficient with the objective vector \(\boldsymbol{y}\) is calculated as:
\begin{equation}\label{PRCC}
    r_{z_j y} = \frac{\operatorname{Cov}\left(\boldsymbol{z_j}, \boldsymbol{y}\right)}{\sqrt{\operatorname{Var}\left(\boldsymbol{z_j}\right) \operatorname{Var}(\boldsymbol{y})}} = \frac{\sum_{i=1}^N \left(z_{ij} - \bar{z}_j\right)\left(y_i - \bar{y}\right)}{\sqrt{\sum_{i=1}^N \left(z_{ij} - \bar{z}_j\right)^2 \sum_{i=1}^N \left(y_i - \bar{y}\right)^2}} \, ,
\end{equation}
where \(\bar{z}_j\) and \(\bar{y}\) denote the mean values of \(\boldsymbol{z_j}\) and \(\boldsymbol{y}\), respectively. 
The PRCC method is particularly suitable for analysing parameters that exhibit monotonic relationships, whether linear or nonlinear, with the selected model outcome.

In our case, two objective functions are employed to evaluate the system, namely the tumour \emph{mass} and \emph{volume}, which are medically relevant indicators of disease progression:
\begin{equation}
M(G(x,y)) = \int_\mathcal{D} G(x,y) \, dx\,dy \, ,
\end{equation}
\begin{equation}
V_{\tau}(G(x,y)) = \int_\mathcal{D} \mathbf{1}_\tau(G(x,y)) \, dx\,dy \, ,
\end{equation}
where \(\mathbf{1}_\tau(G(x,y))\) is a threshold function defined as:
\begin{equation}
\mathbf{1}_\tau(G(x,y)) = 
\begin{cases}
1, & G(x,y) \geq \tau \, ,\\
0, & G(x,y) < \tau \, ,
\end{cases} 
\end{equation}
where the parameter \(\tau\) denotes a threshold value, introduced to identify the effective tumour region, since no sharp interface exists between tumour and healthy tissue in a diffusive concentration model.
The area where the tumour cell concentration \( G \) exceeds this threshold is taken to define the spatial extent of the tumour. Consistently with previous studies \cite{Swanson:2002}, a threshold of 8000 cells/mL is adopted. Given a carrying capacity of \( G_{m} = 2.39 \cdot 10^8 \) cells/mL \cite{Basanta:2011}, this corresponds to a dimensionless concentration threshold of 0.0335. 

\begin{figure}[ht]
\centering
  \includegraphics[width=0.49\linewidth]{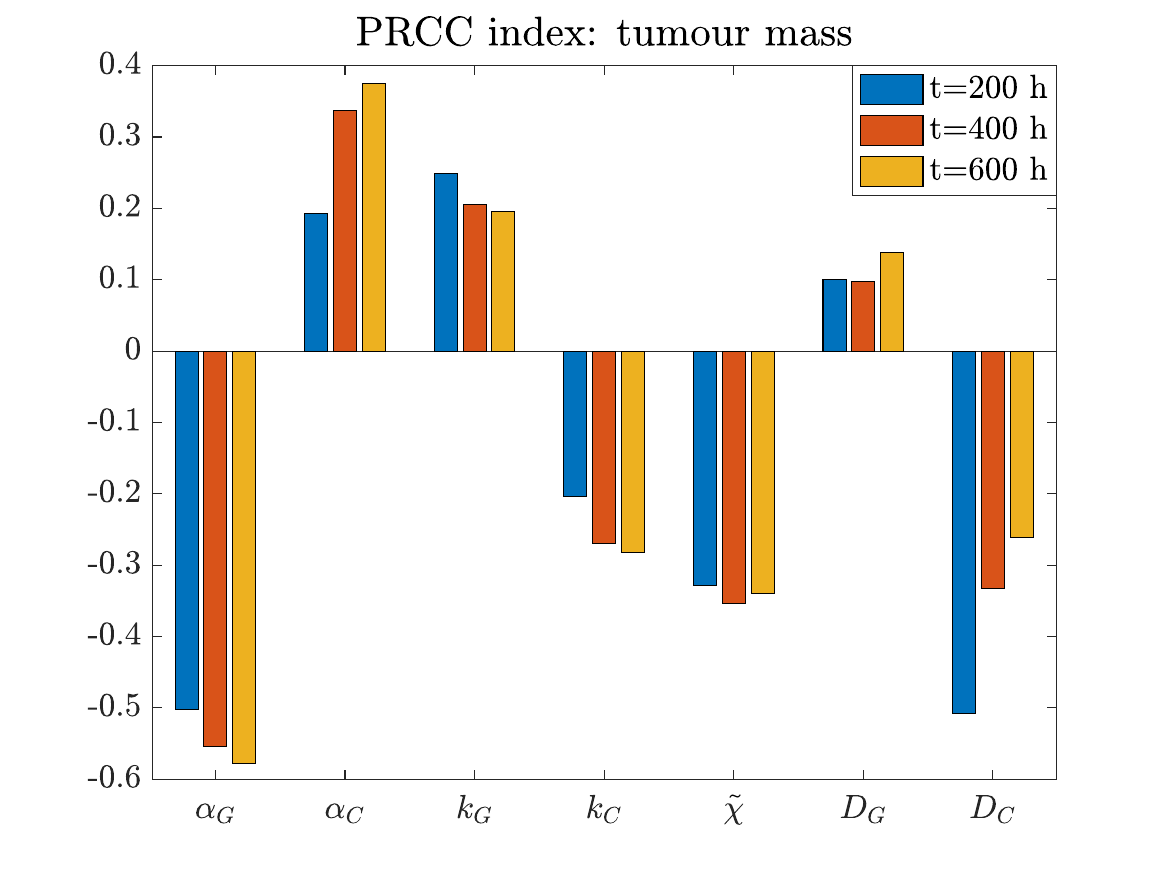} \ 
\includegraphics[width=0.49\linewidth]{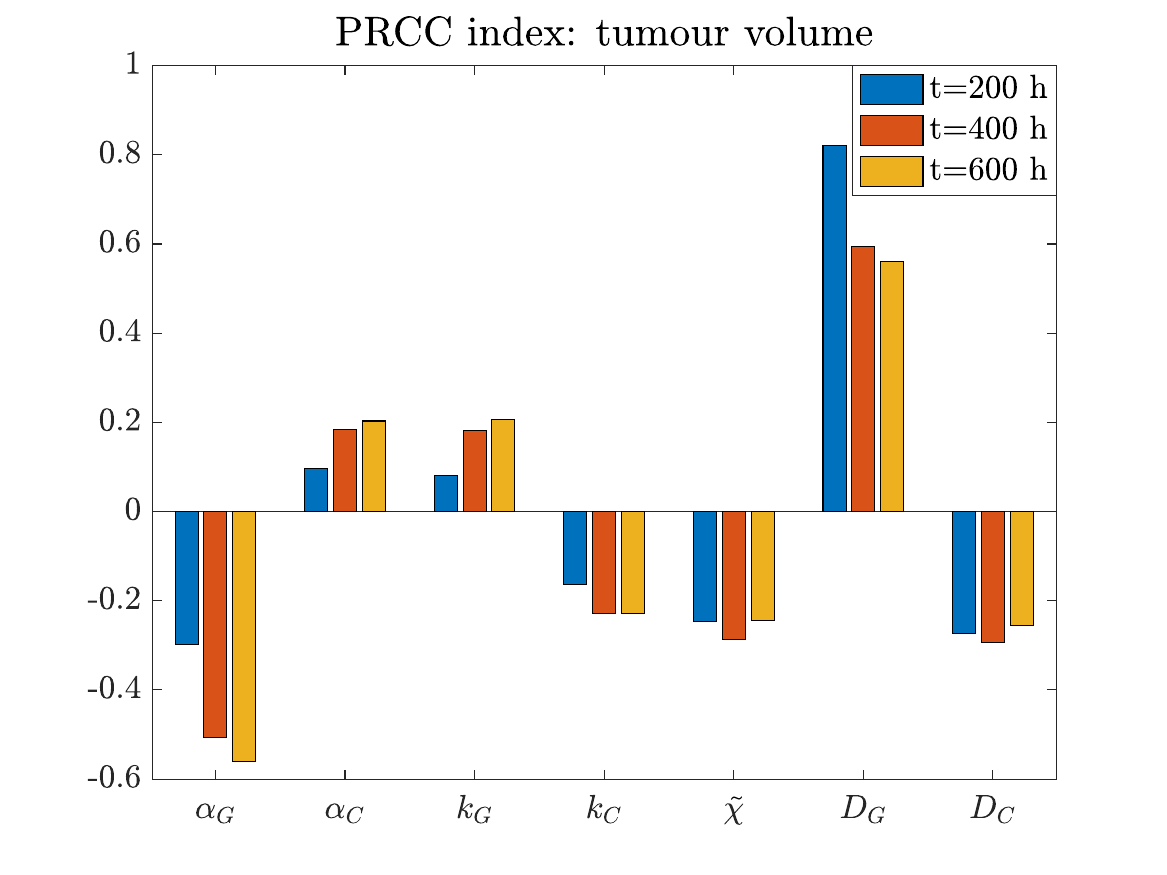}
  \caption{Partial rank correlation coefficient (PRCC) analysis of tumour mass (left) and tumour volume (right) as objective functions.}
  \label{fig:PRCC}
\end{figure}

The sensitivity analysis investigates the effects of several key parameters, including those involved in the Michaelis–Menten kinetics (\(\alpha_G\), \(\alpha_C\), \(k_G\), \(k_C\)), the chemotactic sensitivity parameter (\(\tilde{\chi}\)), and the diffusion coefficients of tumour cells and lymphocytes (\(D_G\) and \(D_C\), respectively).
All parameters are varied within the range of 0.1 to 1.5 times their respective reference values, used in the numerical simulations of Section~\ref{chemio_2D} and reported in Table~\ref{tab:parameters}. The objective functions are computed based on the tumour cell density \(G(x, y)\) at three specific time points, namely \(t = 200\), \(400\), and \(600 \, \text{h}\), to observe both early and late dynamics in tumour progression and immune response. Figures~\ref{fig:PRCC} present the results of the sensitivity analysis, obtained performing a total of \(N = 1200\) simulations.

As expected, the parameter \( \alpha_G \) exhibits a strong negative correlation both with tumour mass, starting from the earliest time points, and with tumour volume, particularly at later stages. This confirms that enhancing the rate at which cytotoxic T-lymphocytes eliminate glioma cells is an effective strategy for reducing tumour burden and limiting its spatial expansion. 

In contrast, the lymphocyte clearance rate \( \alpha_C \), which quantifies the elimination of T-lymphocytes by tumour cells, shows a positive correlation, especially with tumour mass. This behaviour highlights that a higher death rate of CTLs due to tumour activity weakens the immune response, thereby facilitating tumour progression and leading to greater tumour mass accumulation, particularly at later time points. On the other hand, the impact of \( \alpha_C \) on tumour spreading appears to be less pronounced.

Regarding the Michaelis–Menten constants, the glioma half-saturation constant \( {k}_G \) shows a positive correlation, especially with tumour mass at early stages. This suggests that higher values of \( {k}_G \) diminish the effectiveness of the immune response, thereby facilitating more pronounced tumour proliferation, particularly during the early stages of the process. The influence of \( {k}_G \) becomes apparent also in the spatial expansion of the tumour, particularly at later times.

On the other hand, the CTL half-saturation constant \( {k}_C \) displays a negative correlation with both tumour mass and volume. This suggests that lower values of \( {k}_C \) increase lymphocyte elimination even at low tumour cell densities, leading to a weaker immune response and promoting tumour progression.

The chemotactic parameter \( \tilde{\chi} \) also exhibits a strong negative correlation, with both tumour mass and volume, indicating that improved chemotactic responsiveness enhances the ability of immune cells to spatially target tumour regions, thereby supporting tumour eradication. This effect is consistently observed across all time points.

Conversely, the tumour cell diffusion coefficient \( D_G \) consistently shows a strong positive correlation with tumour volume, reflecting how increased tumour motility promotes spatial invasion. A mild positive correlation is also observed between \( D_G \) and tumour mass. 

In contrast, the lymphocyte diffusion coefficient \( D_C \) displays a strong negative correlation, particularly with tumour mass at early time points, suggesting that the initial dispersion of immune cells enhances their spatial coverage of the tumour region and thus improves their cytotoxic effect. The impact of increasing \( D_C \) on reducing the spatial extension of the tumour is less pronounced.

In summary, at earlier times, the parameters \( \alpha_G \), \( D_C \), and \( D_G \) emerge as the most influential in shaping tumour–immune dynamics, as indicated by their high absolute PRCC values. Specifically, the first two primarily control tumour mass, while the latter governs tumour spatial extension. As time progresses, although \( \alpha_G \) and \( D_G \) continue to exert a strong influence, the effect of \( D_C \) becomes less pronounced and other key parameters, such as \( \alpha_C \) and \( \tilde{\chi} \), increasingly contribute to the system's behaviour.
This sensitivity analysis supports the biological plausibility of the model and highlights the importance of accurate parameter calibration to ensure reliable simulation outcomes.

\subsection{Simulations in a 3D brain geometry} 

It is well recognized that the progression of brain tumours is strongly influenced by the surrounding microenvironment \cite{Nóbrega:2024}. In highly anisotropic and heterogeneous tissues such as the brain, tumour evolution is also shaped by the alignment of fibres and the varying properties of different tissue regions. These structural features can vary significantly between individuals and may evolve over time due to ageing or disease. Consequently, patient-specific predictions require patient-specific data.


To this end, we solve the system~\eqref{dimentional_PDE_model} numerically within a brain domain reconstructed from magnetic resonance imaging (MRI) data acquired from patients at the Istituto Neurologico Carlo Besta in Milan (Italy). All imaging data were anonymised, and informed consent was obtained for research purposes. To capture the local anisotropy and fibre orientation, we also incorporate diffusion tensor imaging (DTI) data.
DTI is an advanced MRI technique that measures the directional dependence of water molecule diffusion in brain tissue. This is accomplished by acquiring multiple diffusion-weighted images (at least six, though typically more are used to improve accuracy) using MRI sequences sensitive to the Brownian motion of water molecules.

In the following, we first provide a detailed description of the numerical setup and then compare simulation results for tumours located in different brain regions and subjected to varying levels of lymphocyte injection.

\subsubsection{Numerical simulation setup}
\label{3D_setup}

\paragraph{Initial conditions and therapeutic input.}

The initial tumour cell concentration is modelled as a Gaussian distribution centred at \((x_G, y_G, z_G)\):  
\begin{equation}
    G(x,y,z,t=0) = G_0 \exp\left[\dfrac{(x-x_G)^2 + (y-y_G)^2 + (z-z_G)^2}{2\sigma_G^2}\right] \, ,
\end{equation}  
with \(\sigma_G = 0.44 \, \text{mm}\) and the maximum concentration \( G_0 = 0.5 \), as in the 2D case. 
As the model behaviour is analysed under conditions of spatially varying anisotropy, tumours located in different brain regions exhibit distinct growth patterns. To assess these differences, the system is solved for multiple initial conditions, i.e., for various choices of \( \left(x_G, y_G, z_G\right) \), each leading to different mesh configurations. 

The initial conditions for the chemoattractant and lymphocyte concentrations are set to zero:  
\begin{equation}\label{CI_3D}
    C(x,y,z,t=0) \equiv 0 \, , \quad T(x,y,z,t=0) \equiv 0 \, .
\end{equation}  

As in the 2D simulation, the forcing term which models lymphocyte infusion is represented by a time-independent Gaussian function:  
\begin{equation}
\label{S_3D}
\hat{S}_T(x,y,z) = S_0 \exp\left(- \dfrac{(x-x_T)^2 + (y-y_T)^2 + (z-z_T)^2}{2\sigma_S^2}\right) \, ,
\end{equation}  
with \(\sigma_S = 0.44 \, \text{mm}\). 
Several infusion points \( (x_T, y_T, z_T) \) are considered, each positioned at a fixed distance of 14 mm from the tumour centre, across all examined tumour locations.


\paragraph{Patient-specific mesh creation.}

The patient-specific computational mesh is generated by processing post-contrast T1-weighted magnetic resonance images from a patient affected by GBM. The initial step involves image segmentation through \textit{3D Slicer}'s automatic segmentation \cite{3D_Slicer}, dividing it into distinct and homogeneous regions, emphasizing the areas of interest to facilitate analysis. The next step involves extracting the brain external surface from the segmented map using \textit{Vmtk} (Vascular Modelling ToolKit) \cite{vmtk}. 
Starting from the gray-scale MRI scans, computational meshes are generated using the open-source software \emph{TetGen}~\cite{Tetgen} to construct a conforming tetrahedral mesh from a list of nodal points, derived from the MRI.
The mesh is selectively refined in the regions designated for tumour growth and therapy to ensure adequate spatial resolution. Specifically, local mesh refinement is achieved by assigning a target edge length \( h \) to each node, which controls the maximum size of the elements adjacent to that point. In the region where the tumour is expected to expand, a refinement value of \( h = 0.6 \, \mathrm{mm} \) is used, while in the region where lymphocyte motion occurs, a finer resolution of \( h = 0.3 \, \mathrm{mm} \) is adopted.

\begin{figure}[ht]
\centering
  \includegraphics[width=\linewidth]{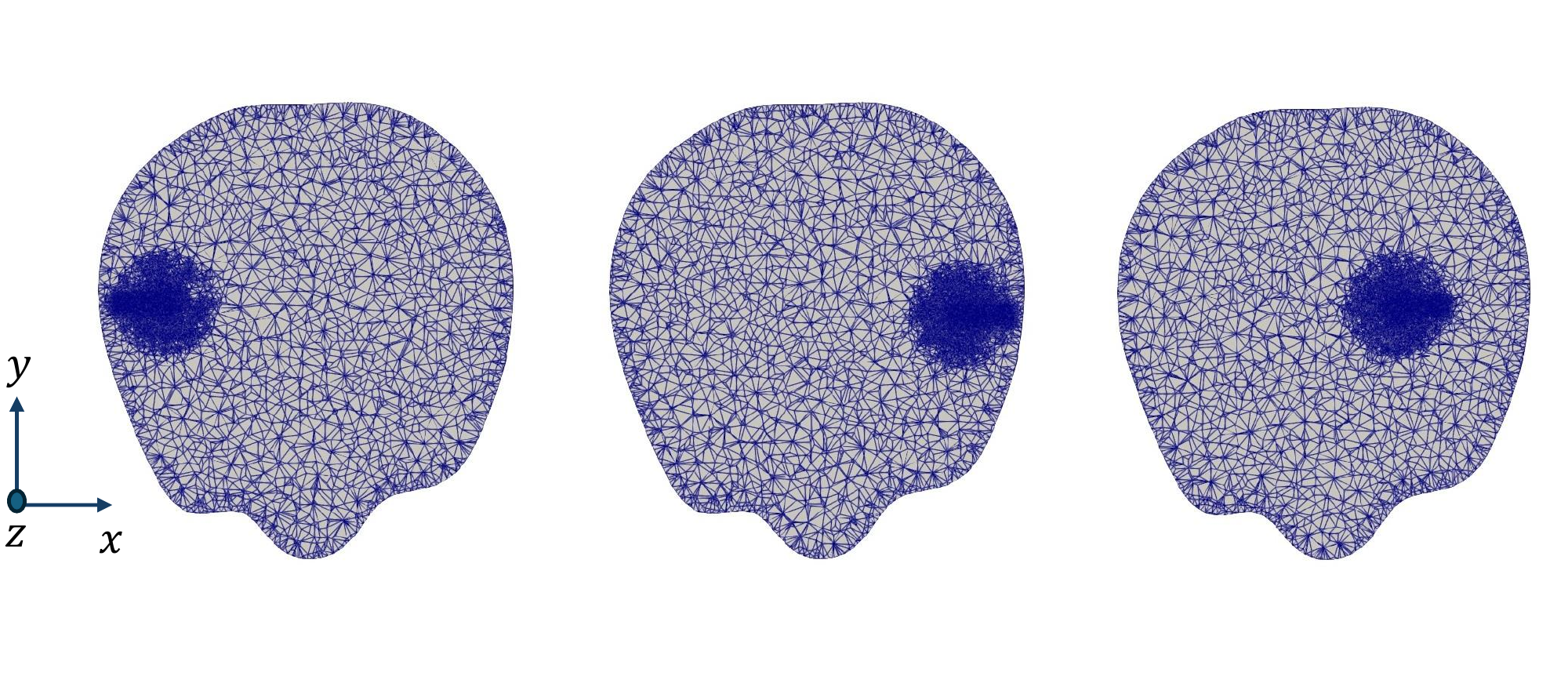}
\caption{
Cross-sectional views of the computational meshes. From left to right, the tumour is positioned in: the right hemisphere \((193\, \text{mm},\, 273\, \text{mm},\, 17\, \text{mm})\), the left hemisphere \((289\, \text{mm},\, 273\, \text{mm},\, 17\, \text{mm})\) and the central-left region \((265\, \text{mm},\, 277\, \text{mm},\, 17\, \text{mm})\).}
\label{fig:mesh_latoSX}
\end{figure}

Three different meshes are constructed, each corresponding to a distinct initial tumour location. Figure~\ref{fig:mesh_latoSX} shows a representative cross-section of all three meshes, obtained by slicing each domain along a plane passing through the tumour centre. From left to right, the tumour locations correspond to the right, left, and central-left hemispheres.

We then account for the fact that brain tissue is composed of highly oriented fibres, such as axonal tracts, which give rise to anisotropy, and that it also includes various tissue types and water compartments, resulting in heterogeneous properties. 
To incorporate these features, we first generate six meshes corresponding to the independent components of the symmetric diffusion tensor derived from the DTI images. In addition, we construct a labelled mesh encoding the spatial distribution of cerebral tissues, which is used to assign heterogeneous motility properties to both cells and signalling molecules.

Specifically, from the patient-specific DTI, a symmetric second-order diffusion tensor \( \mathbb{D}_w \) is reconstructed, representing the local diffusivity profile of water in three dimensions. 


Then, since glioma cells and lymphocytes tend to migrate preferentially along the same directions revealed by DTI for water diffusion, 
from the tensor \( \mathbb{D}_w \) we define the tensor $\mathbb{D}_{p d}$ representing the preferential directions of motion.
This tensor is constructed by enhancing mobility along the principal diffusion directions identified by DTI, following the methodology proposed in~\cite{Jbabdi:2005, Lucci:2022, Ballatore:2023, Ballatore:2024}. This strategy has been shown to improve the accuracy of tumour growth modelling by better capturing the directional nature of cell migration~\cite{Jbabdi:2005}.
To this end, we define the linear \(c_l\), planar \(c_p\), and spherical \(c_s\) anisotropy indices:
\begin{equation}
\begin{aligned}
c_l & =\frac{\lambda_1-\lambda_2}{\lambda_1+\lambda_2+\lambda_3} \, , \quad
c_{\mathrm{p}} & =\frac{2\left(\lambda_2-\lambda_3\right)}{\lambda_1+\lambda_2+\lambda_3} \, , \quad
c_{\mathrm{s}} & =\frac{3 \lambda_3}{\lambda_1+\lambda_2+\lambda_3} \, ,
\end{aligned}
\end{equation}
where \(\lambda_1\), \(\lambda_2\), and \(\lambda_3\) are the eigenvalues of the tensor \(\mathbb{D}_w\), ordered by magnitude. The tensor of preferential directions $\mathbb{D}_{p d}$ is then constructed as
\begin{equation}
\mathbb{D}_{p d}=a_1(r) \lambda_1 \hat{e}_1 \otimes \hat{e}_1+a_2(r) \lambda_2 \hat{e}_2 \otimes \hat{e}_2+ \lambda_3 \hat{e}_3 \otimes \hat{e}_3 \, ,
\end{equation}
where \(\hat{e}_i\) are the eigenvectors of \(\mathbb{D}_w\) associated with \(\lambda_i\), and \(a_i(r)\) are coefficients dependent on a tuning parameter \(r\), defined as
\begin{equation}
\label{a_i}
\begin{aligned}
& a_1(r)=r\left(c_l+c_{\mathrm{p}}\right)+c_{\mathrm{s}} \, ,\\
& a_2(r)=c_l+r c_{\mathrm{p}}+c_{\mathrm{s}} \, .
\end{aligned}
\end{equation}
This approach allows for the following adjustments: when one eigenvalue dominates (i.e. \(c_l \sim 1\)), the diffusion is increased along the principal direction; in the case of fibre intersections along two different directions (\(c_p \sim 1\)), diffusion is restricted in the direction orthogonal to the plane spanned by the two principal eigenvectors; and when no preferential direction is present (\(c_s \sim 1\)), the tensor remains unchanged.

Moreover, 
the brain is not a homogeneous medium: it comprises grey matter and white matter, while the ventricles and other cavities surrounding the cerebellum are filled with cerebrospinal fluid. These different media affect cell motility differently \cite{Swanson:2002}. 

\begin{figure}[ht]
\includegraphics[width=\linewidth]{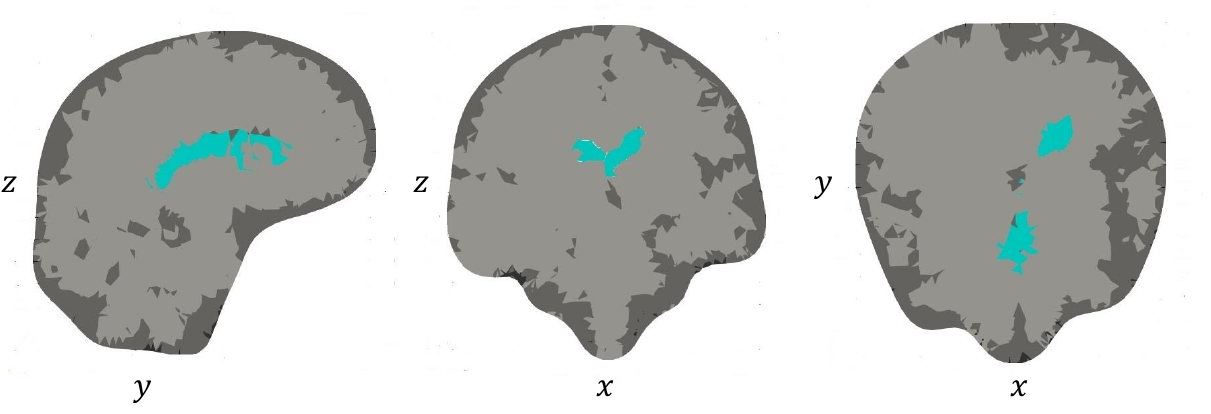}
\caption{Segmentation: sagittal, coronal, and transverse sections.}
\label{fig: Segmentazione}
\end{figure}

Using the \emph{Slicer} software \cite{3D_Slicer}, the T1-weighted MRI images were segmented to classify each cell in the computational domain according to the type of brain tissue it represents. The segmentation distinguishes three categories: grey matter, white matter, and cerebrospinal fluid. Figure~\ref{fig: Segmentazione} shows the segmented brain across three orthogonal planes. Dark grey areas correspond to grey matter, light grey areas to white matter, and light blue regions identify the ventricles filled with cerebrospinal fluid.
 

We note that the automated segmentation failed to correctly distinguish anatomical structures deep within the right hemisphere (see the third panel of Fig.~\ref{fig: Segmentazione}), due to the presence of a tumour in the patient from whom the MRI data were acquired. To ensure the reliability of the numerical simulations, this non-segmented region will not be selected as the initial tumour site. 

Finally, to obtain the diffusion tensors for the various populations, the tensor of preferential directions is normalised by dividing it by one-third of its trace, thereby preserving its anisotropic structure  while rescaling the magnitude so that the average diffusivity equals one.
 The resulting dimensionless tensor is then multiplied by the diffusion coefficient specific to each species, yielding the final anisotropic diffusion tensors used in the simulations:
\begin{equation}
\begin{aligned}
   \mathbb{D}_G&=D_G\dfrac{3\mathbb{D}_{pd}}{\text{tr}(\mathbb{D}_{pd})} \, , \quad 
   \mathbb{D}_C&=D_C\dfrac{3\mathbb{D}_{pd}}{\text{tr}(\mathbb{D}_{pd})} \, , \quad 
   \mathbb{D}_T&=D_T\dfrac{3\mathbb{D}_{pd}}{\text{tr}(\mathbb{D}_{pd})} \, .
\end{aligned}
\end{equation}

The values of \( D_G \), \( D_C \), and \( D_T \) depend on the tissue type. Notably, tumour cells diffuse approximately five times faster in white matter than in grey matter~\cite{Swanson:2000}. In regions filled with cerebrospinal fluid, cells are assumed to be non-motile, as they generally require adhesion to a substrate to migrate. Moreover, it is assumed that chemoattractants do not cross the barrier formed by the ependymal cells lining the ventricular walls and, therefore, cannot diffuse within these regions.

Therefore, the diffusion coefficients are assigned as follows:
\begin{itemize}
    \item in voxels corresponding to white matter, \( D_G \), \( D_C \), and \( D_T \) take the reference values reported in Table~\ref{tab:parameters};
    \item in voxels corresponding to grey matter, \( D_G \), \( D_C \), and \( D_T \) are scaled by a factor of 0.2, in accordance with estimates available in the literature~\cite{Swanson:2000,Swanson:2002};
    \item in voxels corresponding to cerebrospinal fluid, all diffusion coefficients are set to values close to zero.
\end{itemize}

\paragraph{Time discretization.}

Finally, as in the 2D setup, the CFL condition is also verified in the 3D simulations. To this end, it is first necessary to estimate the chemotactic coefficient, as it determines the maximal expected lymphocyte velocity, which in turn is used to compute the Courant number.

A simplified simulation is thus performed by solving the system for \(C\) and \(T\) only, with \(\hat{b}_C = 0\) and \(\hat{S}_T(x,y,z) \equiv 0\), following the same approach used in the 2D case. Results show that at \(t = 1200 \, \text{h}\), the maximum gradient of the chemoattractant reaches \(\overline{\nabla T} = 0.0144 \, \text{mm}^{-1}\). Assuming, as before, a reference lymphocyte velocity of \(v = 0.24 \, \text{mm/h}\) \cite{Letendre:2015}, the chemotactic coefficient is estimated as \(\tilde{\chi} = 16.6 \, \text{mm}^2 \, \text{h}^{-1}\). The same simulation also provides the value of the chemoattractant outside the tumour at \(t = 1200 \, \text{h}\). Assuming a 30\% reduction in lymphocyte speed inside the tumour, the saturation parameter is set to \(\epsilon = 1.6\). 

Under these assumptions, the expected lymphocyte speed (neglecting anisotropy) is \( u \leq 0.24 \, \text{mm/h} \). The spatial discretisation is estimated using the smallest edge length of the mesh elements, set in \emph{TetGen} as \( h_{\text{min}} = 0.3 \, \text{mm} \). A time step of \( \Delta t = 0.167 \, \text{h} \) is adopted, resulting in an estimated Courant number of \( C_T \sim 0.133 \), which satisfies the stability condition \( C_T < 1 \). This conservative choice guarantees numerical stability even in the presence of anisotropy, where the scheme is more sensitive. Additionally, a runtime check in the numerical implementation ensures that the CFL condition is never violated during the simulations.

\subsubsection{Numerical simulation with the patient-specific data} 

\begin{figure}
 \includegraphics[width=0.9\linewidth]{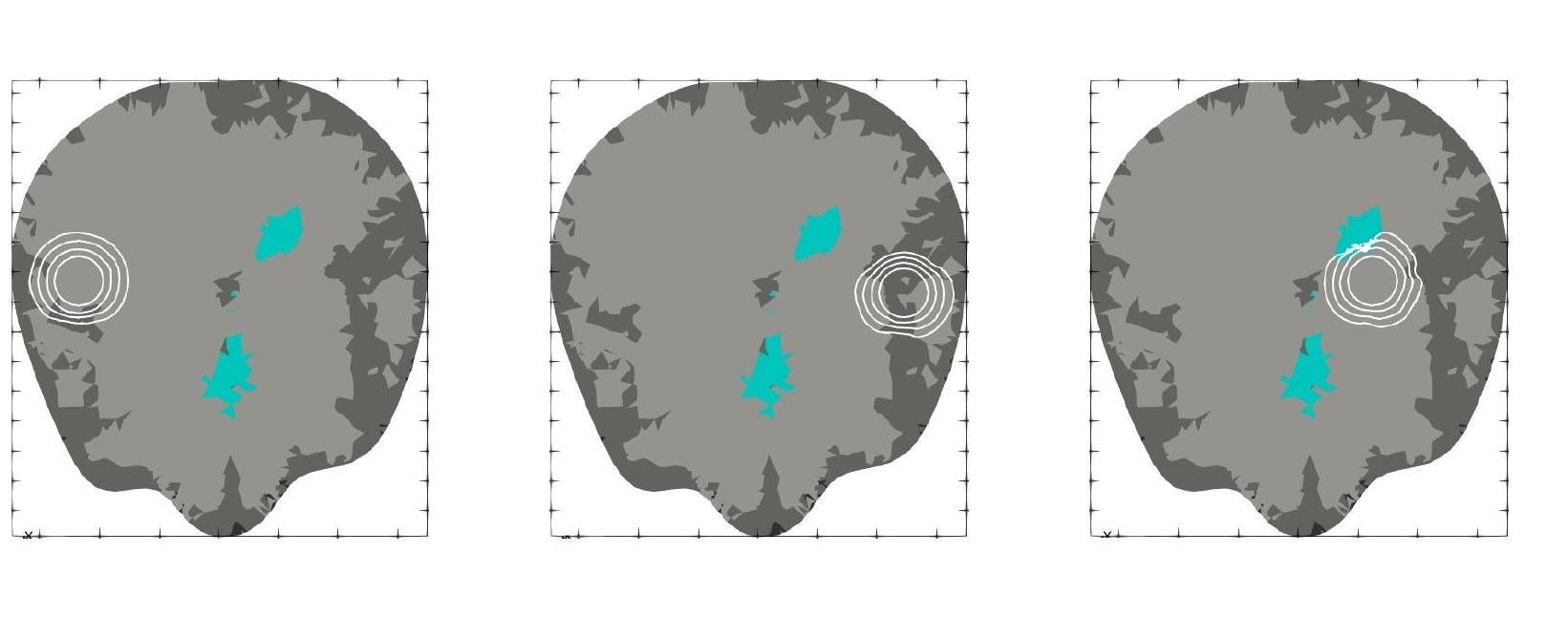} \\
 \includegraphics[width=0.9\linewidth]{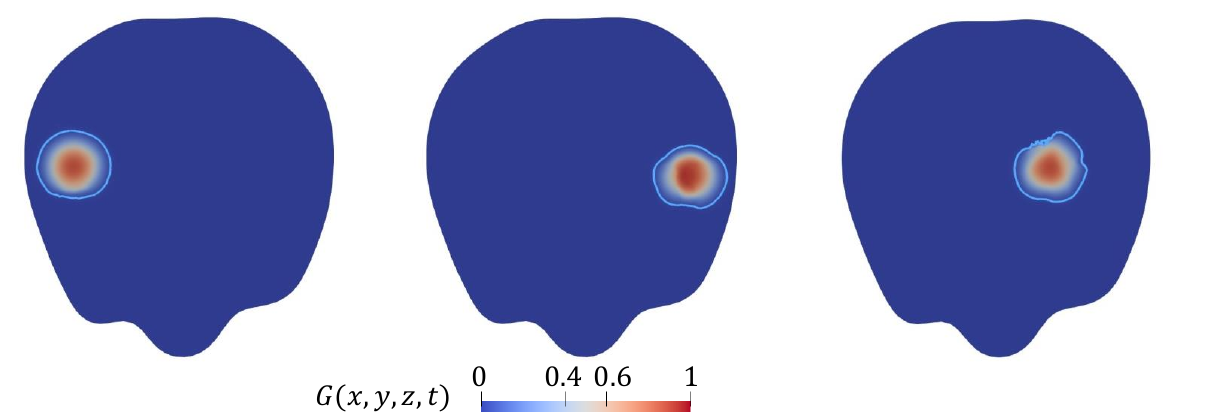}
\caption{First row: tumour growth profiles in the right, left, and central-left hemispheres at $t = 0, 400, 800,$ and $1200 \,\mathrm{h}$. 
Second row: tumour cell concentration in the same hemispheres at $t = 1200 \,\mathrm{h}$. 
The contour line marks the region where the tumour concentration exceeds $0.0335$, delineating the tumour-occupied zone.
}
\label{fig:contour}
\end{figure}

The system of equations \eqref{dimentional_PDE_model} is first numerically solved in the absence of therapy, considering different initial tumour locations as shown in Fig.~\ref{fig:mesh_latoSX}. The diffusion tensor accounts for both DTI and tissue-type data, with an anisotropy amplification factor set to \( r = 5 \) in Eqs.~\eqref{a_i}. Fig.~\ref{fig:contour} shows the growth profiles of the tumour for different initial locations at $t = 0, 400, 800,$ and $1200 \,\mathrm{h}$, plotted on the segmentation of a transverse section. Moreover, the second row illustrates the tumour growth concentration at \( t = 1200 \, \mathrm{h} \). 

The results reveal significant deviations from the spherical growth pattern expected in the isotropic case. Tumour expansion is notably hindered in regions of grey matter and cerebrospinal fluid, underscoring the role of brain tissue heterogeneity in shaping tumour progression. Furthermore, the tumour expands preferentially along the directions indicated by the DTI data, resulting in an elongated shape aligned with the underlying fibre structures.
The effects of anisotropy and heterogeneity are particularly evident in the left and central-left ventricles, whereas they are less pronounced in the right hemisphere due to segmentation failure in the region affected by the patient's tumour. In that area, the tumour disrupts fibre structures and causes fluid accumulation, resulting in a more isotropic environment.

To enable meaningful comparisons with clinical data, we computed several tumour growth metrics commonly used in biomedical research. After 1200 hours, the tumour reaches volumes of \(15095.8 \text{ mm}^3\), \(13121.3 \text{ mm}^3\), and \(14961 \text{ mm}^3\) in the right, left, and central-left hemispheres, respectively, demonstrating substantial growth in all scenarios. Notably, we estimated a volume doubling time (VDT) ranging from 17 to 18.5 days, which closely aligns with values reported in clinical studies. For reference, Ellingson et al. \cite{Ellingson:2016} found a median VDT of 21.1 days. Additionally, we computed the specific growth rate (SGR), defined as \( (\ln 2)/\text{VDT} \), obtaining a value of approximately \(4\%\) per day, which is also consistent with the mentioned references. Another commonly used metric for assessing brain tumour progression is the average radial expansion velocity (VRE). In our simulations, we obtained \( v_{\mathrm{RE}} = 6.24 \cdot 10^{-4} \, \mathrm{cm/h} \), \( v_{\mathrm{RE}} = 5.66 \cdot 10^{-4} \, \mathrm{cm/h} \), and \( v_{\mathrm{RE}} = 6.20 \cdot 10^{-4} \, \mathrm{cm/h} \) in the three different scenarios, respectively. 
These findings indicate that the simulation outcomes closely replicate the natural progression of the disease. Indeed, glioblastomas are known to reach diameters of up to \(6\, \mathrm{cm}\) before leading to patient death~\cite{Swanson:2002}, with reported radial expansion rates of up to \(v_{\mathrm{RE}} = 7.5 \cdot 10^{-4}\, \mathrm{cm/h}\)~\cite{Swanson:2000}. 
For the tumour initially located in the left hemisphere, which represents the most anisotropic scenario, it is worth computing both the maximal and minimal velocities of radial expansion. The maximal radial expansion velocity is $v_{\mathrm{RE}}^M = 7.27 \cdot 10^{-4}\, \mathrm{cm/h}$, while the minimal velocity is $v_{\mathrm{RE}}^m = 4.28 \cdot 10^{-4}\, \mathrm{cm/h}$, highlighting a substantial difference between the two.

\begin{figure}[ht]
\includegraphics[width=\linewidth]{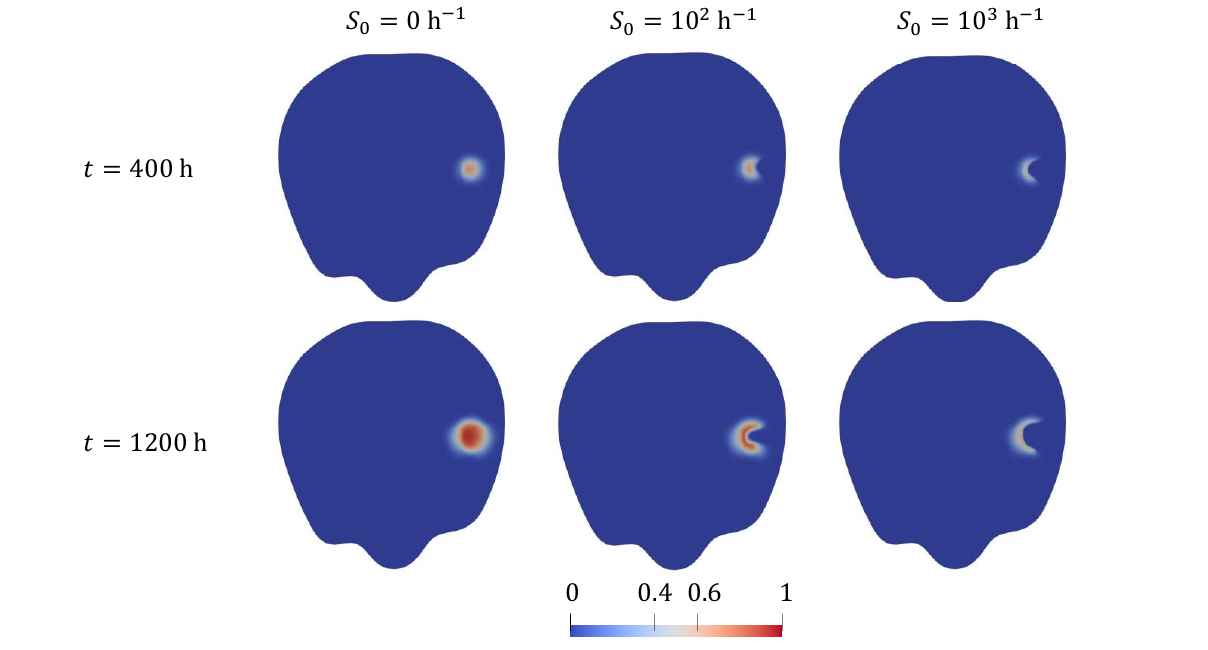}
\caption{Tumour progression in the left hemisphere at various time points and under different therapy intensities. }
\label{fig:growth_with_therapy_sx}
\end{figure}

Subsequently, therapy is introduced to counteract tumour progression. 
Figure~\ref{fig:growth_with_therapy_sx} illustrate the therapeutic effects on a tumour located in the left hemisphere, which is the one in which the effect of anistropicity are more evident. 
Results are reported in terms of tumour concentrations at \( t = 400 \, \mathrm{h} \) and \( t = 1200 \, \mathrm{h} \), comparing cases without therapy and with different rate of lymphocyte infusion,  \( S_0 \),  rescaled with respect to \( C_m \). 
 The analysed scenarios correspond to total lymphocyte infusion rates of \( 0 \, \text{cells/h} \) (i.e., no therapy), \( 1.34 \cdot 10^5 \, \text{cells/h} \), and \( 1.34 \cdot 10^6 \, \text{cells/h} \). 

The results confirm that increasing the therapy intensity enhances treatment efficacy, achieving effective tumour control at the highest infusion rate.
Although therapy does not fully halt tumour growth in all cases, it significantly slows its progression, highlighting its potential as a modulatory tool when appropriately dosed and timed.
Notably, the highest therapy value considered are below the infusion rates commonly explored in clinical research, where up to \( 10^7 \) lymphocytes (e.g., CAR-T cells) are administered in a single day and often repeated on a weekly basis~\cite{Brown:2016}.

\begin{figure}[ht]
\includegraphics[width=\linewidth]{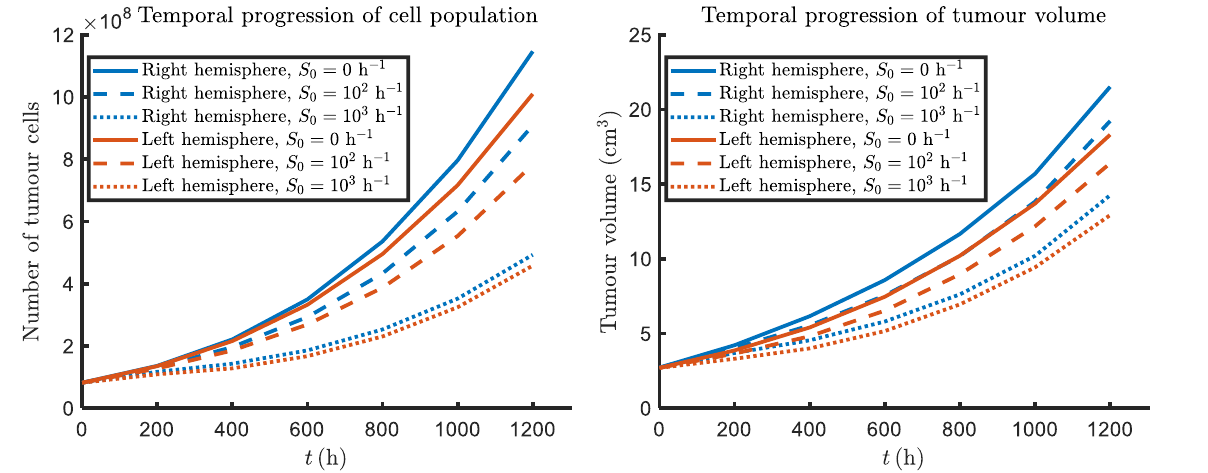}
\caption{Therapeutic impact on tumour cell population and volume for different tumour location and varying the rescaled therapy intensities: \( S_0 = 0 , \,  10^2 , \, 10^3 \ {\rm h}^{-1}\).}
\label{fig:ani_matlab}
\end{figure}

Furthermore, we remark that some slightly differences in tumour progression and therapy outcomes are perceivable, depending on the location in which the tumour is first localized.
The two graphs in Figure \ref{fig:ani_matlab} summarize the temporal evolution of the tumour cell population and tumour volume for different therapy intensities and tumour locations. Despite the application of therapy with an intensity of \( S_0 = 10^3 \ {\rm h}^{-1}\), the treatment is unable to completely halt tumour growth but effectively slows its progression, reducing the cell population by half, or in the right hemisphere to as little as one-third. A similar reduction is observed in tumour volume, with therapy resulting in a volume that is three-quarters of that in the untreated case.

The results also reveal notable differences in tumour growth depending on the surrounding tissue type. Specifically, the tumour in the left hemisphere, which is surrounded by regions with a higher proportion of white matter, exhibits a volume approximately 17.52\% larger and a 13.55\% higher cell population at $t = 1200 \, \mathrm{h}$ compared to the tumour in the right hemisphere in the absence of therapy. These findings highlight the influence of tissue heterogeneity on tumour dynamics. It should be noted that these differences are substantially reduced under the maximal therapy scenario.






\section{Conclusions and future developments}

Glioblastoma Multiforme represents one of the most lethal and aggressive forms of cancer that develop in the brain. Despite advancements in medicine, its prognosis remains predominantly poor, motivating the exploration of novel therapeutic approaches. In this context, immunotherapy emerges as a promising strategy.

This work aims to develop mathematical models that can describe the complex tumour-immune interactions. The proposed model, formulated using PDEs, focuses on the interactions between tumour cells and lymphocytes to represent a specific type of immunotherapy: the infusion of cytotoxic T lymphocytes, a therapeutic approach that is the subject of extensive research globally.

As a first approximation, disregarding the spatial distribution of cells, the PDE model is simplified into an ODE system consisting of two equations: one representing the population of tumour cells and the other the number of lymphocytes in that region, with therapeutic infusion modelled as a time-constant source term in the second equation. The system's behaviour is studied under varying therapy intensities. Analytically, a threshold is derived that guarantees the system evolves towards a healthy equilibrium, where the tumour cell population is eradicated. The corresponding lymphocyte infusion levels are comparable to those observed in clinical studies. Numerical methods are employed to generate bifurcation diagrams of the system, exploring variations in each parameter.

The PDE model expands the analysis by incorporating the spatial dynamics of tumour progression. The diffusive terms account for the spatial expansion of the cells. Additionally, a chemotactic term is included in the lymphocyte equation to model their directed movement towards tumour-secreted chemical signals. Various modelling approaches for the chemoattractant are also explored and discussed. This model is solved using a finite element code implemented in Python, leveraging the FEniCS libraries. 
Initial simulations are performed on a simplified two-dimensional geometry, specifically a circular sector, to test code stability and model behaviour. These simulations illustrate glioma proliferation and spatial expansion while lymphocytes diffuse and migrate via chemotaxis to attack tumour cells. Based on these two-dimensional simulations, a sensitivity analysis is conducted.
The parameters with the greatest impact on tumour mass are the cancer cell elimination rate by lymphocytes and lymphocyte diffusivity, especially at early times. Over longer time scales, additional factors such as the lymphocyte elimination rate and the chemotactic coefficient also significantly influence tumour mass. Regarding tumour volume, a critical prognostic indicator, the most influential parameter is the glioma cell diffusion coefficient, while at later stages the elimination rate of glioma cell mediated by lymphocyte also becomes relevant.
These findings indicate that these parameters should be considered key targets in the development of synergistic therapeutic strategies.

The system is subsequently simulated in a three-dimensional brain geometry reconstructed from MRI data of a patient. Brain structure can vary significantly between individuals and may change over time due to aging or disease. These variations affect both tumour evolution and treatment response. Therefore, to enable patient-specific predictions, personalised data are essential. In particular, diffusion tensor imaging (DTI) data are used to account for anisotropy and improve predictions of tumour-invaded regions. This approach is based on the hypothesis that both cells and fluids in the brain preferentially migrate along the spatial arrangement of tissue fibres. The brain is further segmented into different tissue types, namely white matter, grey matter, and cerebrospinal fluid, and the diffusion tensor is appropriately modified for each voxel's composition.
Therapeutic infusion is assumed to be delivered directly into the brain, as described in medical literature, and is modelled as a time-constant and spatially Gaussian source term in the lymphocyte equation.  Under these conditions, the therapy's effects are evaluated for varying intensities and different tumour initial locations.  Results are presented for simulations under this setup, including both free tumour growth in the absence of therapy and the response to varying infusion intensities. Differences are observed based on the tumour's growth location. In all cases, therapy successfully slows tumour growth to varying degrees depending on the lymphocyte infusion rate. However, it does not achieve complete tumour eradication. 

Multiple directions for further research remain open. In this work, only time-constant infusions are considered. However, in clinical practice, therapy administration often occurs in intervals spanning several days. The study of time-varying therapy regimes could be undertaken using the ODE model to identify treatment schedules that minimise intervention while reducing adverse effects, drawing on control theory principles. The PDE model could also be refined by considering alternative infusion techniques. For improved numerical stability, stabilisation techniques such as the streamline upwind Petrov–Galerkin (SUPG) formulation might be employed. This analysis focuses exclusively on interactions between lymphocytes and tumour cells, neglecting contributions from other cell populations (e.g. macrophages and stem cells) as well as cytokines mediating intercellular interactions, which could be incorporated into the model in future work. Finally, exploring the interaction between immunotherapy and other treatment modalities, such as radiotherapy and chemotherapy, may pave the way for developing multimodal strategies.

\section*{Author contributions}
F.B.: formal analysis, conceptualization, investigation, methodology, software, visualization, writing-original draft, writing-review and editing; \\
L.S.: formal analysis, conceptualization, investigation, methodology, software, visualization, writing-original draft; \\
C.G.: formal analysis, conceptualization, investigation, methodology, writing-review and editing, supervision; \\
All authors gave final approval for publication and agreed to be held accountable for the work performed therein.

\section*{Acknowledgments}
F.B and C.G. conducted the research according to the inspiring scientific principles of the national Italian mathematics association Indam (“Istituto nazionale di Alta Matematica”), GNFM group. FB acknowledges support from the PNRR M4C2 through the project "Made in Italy Circolare e Sostenibile (MICS)", CUP: E13C22001900001.
CG acknowledges support from the Italian Ministry of University and Research (MUR) through the grant PRIN2022-PNRR project (No. P2022Z7ZAJ, CUP: E53D23018070001).

\printbibliography[heading=bibintoc]
\end{document}